\documentclass[aps,prl,preprint,groupedaddress]{revtex4-1}
\usepackage{bbm}
\usepackage{graphicx}% Include figure files
\usepackage{subfigure}
\usepackage{amsmath}
\begin{document}

\title{Quantum cellular automata for word statistics facilitated by quantum correlations}

\author{Guanhua Chen$^{1}$ and Yao Yao$^{1,2}$\footnote{Electronic address:~\url{yaoyao2016@scut.edu.cn}}}

\address{$^1$ Department of Physics, South China University of Technology, Guangzhou 510640, China\\
$^2$ State Key Laboratory of Luminescent Materials and Devices, South China University of Technology, Guangzhou 510640, China}

\date{\today}

\begin{abstract}
We propose an iterative algorithm to investigate the cooperative evolution dominated by information encoded within state spaces in a random quantum cellular automaton. Inspired by the 2-gram model in statistical linguistics, the updates of quantum states are determined by a given corpus, which serves as the interactions to induce quantum correlations. These local cooperative interactions lead to block-diagonal evolution quantified by the entanglement asymmetry. We evaluate the influence of two-site gates on the iteration process and reveal the associated search speedup. Crucially, we demonstrate the information scrambling dependent of gate sampling, uncovering an intriguing bond percolation. Our results provide an adaptive paradigm for quantum search algorithms and random many-body dynamics.
\end{abstract}

\maketitle

\textit{Introduction.--} In many-body systems particularly experimentally realizable quantum simulators such as superconducting circuits \cite{2019prlscentanglement,2023nat51entanglement}, the emergence of quantum correlation typically relies on the complex local interactions among components \cite{2016jpacorrelation,2023RPPcorrelation,2015natcorrelation,2016scicorrelation}. This has led to groundbreaking progresses in the field of quantum information, including information scrambling \cite{2022prlscrambling,2022prrscrambling}, quantum thermalization \cite{2016scithermalization}, and long-range entanglement \cite{2020prrmultientanglement,2021scilongentanglement}, which inspire the practical quantum computing with manipulable resources. However, a clear relation between interactions and quantum correlation is still lacking due to the diversity and complexity of Hamiltonian formalism.

Recently, a wide range of algorithms and models developed in natural language processing (NLP), due to the ability to deal with complex systems with multiple degrees of freedom \cite{2015scinlp,2023aipnlp,2022asqnlp}, have been increasingly applied to the study of quantum many-body systems \cite{2018qnlp,2023IEEEnlp}, including recurrent neural networks \cite{2020prrnw} and transformer architectures \cite{2023prbTransformer,2023prlTransformer}, utilized to effectively compute the ground-state wave functions. It has also been proposed that the statistical mechanics can be employed to characterize the probabilities associated with word occurrences and construct pairwise interactions \cite{2010presml,2020entropysml}. In this context, we try to induce quantum correlation between words by the most fundamental statistical language model: $n$-gram model that the occurrence of a word is merely related to the $n-1$ words before it \cite{2024bookn-gram,2014bookn-gram,2019IEEEn-gram,2019esan-gram}, effectively constituting a Markov chain of words, which offers a way to build local cooperative interactions in a particular order.

\begin{figure}[htbp]
	\includegraphics[scale=1]{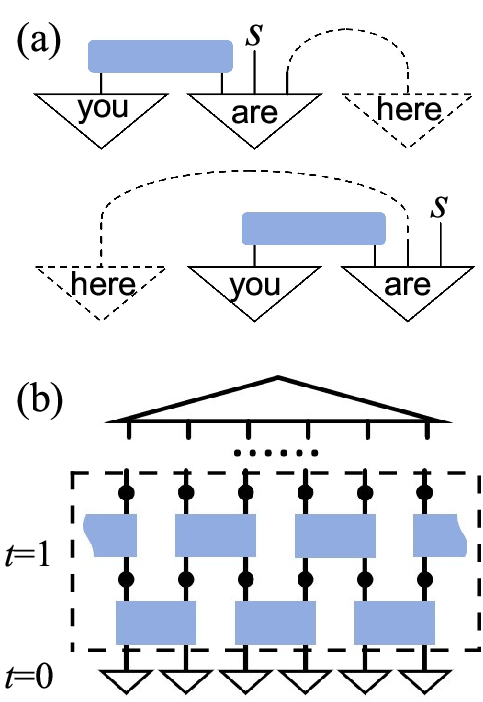}
	\caption{\label{fig1}  (a) Sentence diagram for `\textsf{You are here}' and `\textsf{Here you are}'. A left $s$ wire verifies the grammaticality. The blue blocks with rounded corners denote words \textsf{you} and \textsf{are} are bound together. (b) Schematic representation of the random unitary circuit used to describe the cellular automaton. Inverted triangles are initial random vectors. Every time step is composed of four-layer gates, alternating two-site gates acting on pairs of neighboring sites (blue blocks) and one-site gates for all sites (black circles). The resulting triangle denotes that quantum correlation exists in the whole system. }
\end{figure}

Quantum speedup search in unordered systems exemplified by Grover algorithm, presents one of the central challenges in quantum computing \cite{1996Grover,1998prlsearch,2017qifsearch,2018IEEEsearch,2020prasearch,2024prrsearch,2010jpasearch}. Unlike the traditional dynamical framework based on Hamiltonian, there is no explicit interaction between quantum states during the search process. So it is natural to consider the question, whether we might introduce interactions into the evolution of quantum search based on the information encoded within state spaces? Furthermore, does such a prototype hold generic physical significance?
This statement has been widely considered in the structure searching \cite{2020npjsearch,2023npjsearch,2023nsrsearch}. For instance, in the optimization of two-dimensional boron-carbon-nitride (BCN) materials, the tendency for B-N bonding arising from special interactions, results in the formation of separated graphene and BN domains after annealing \cite{2010nmbcn,2013ncbcn,2016pccpbcn}, which also manifests the cooperative interactions between components fundamentally affect the evolution. Moreover, owing to the spatial correlation (absent in the Gaussian universality class) generated by the interaction of adjacent bonds, there is a ballistic-deposition growth in the surface growth model for entanglement of the Kardar-Parisi-Zhang (KPZ) universality class \cite{1986prakpz,1993prlkpz,2017prxkpz,2011arxivkpz}. In this regard, it is intriguing to construct a cooperative interaction between quantum states and probe corresponding evolution during the search.

In this Letter, we propose an iterative search algorithm with postselection, designed for a random cellular automaton derived from the 2-gram model that the selection of a word is solely influenced by its nearest neighbor, without considering more distant context. The cooperative evolution of quantum states is entirely governed by the characteristics of a given corpus which plays the role of interaction among sites. We compare the random circuits with or without two-site gates and reveal a bond percolation feature depending on effective gates. Labeled by the entanglement asymmetry, local cooperative interactions induce block diagonalization in the squeezed phase space. Furthermore, our analysis specifically focuses on the intricate relation between quantum correlation and the distribution of effective gates, depicting the emergent information scrambling with percolation threshold preceding the critical point of iteration for two-site gates. That is, notably, the spreading of information needs more bonds.

\textit{Setup.--} Our starting point is to build the $2$-gram model in a quantum version to replace conventional two-body interactions: Namely the presence of a state is directly determined by states of the nearest sites, called the local cooperative interaction. The first step is to define the Corpus State ($|{\rm CS}\rangle$) according to a corpus with grammatically correct sentences. Associating the combination of words with a one-dimensional chain of $L$ sites, the total Hilbert space is the tensor product of the spaces corresponding to each site, $\mathcal{H}=\otimes_{j\in L}\mathcal{H}_j$, where dim$(\mathcal{H}_j)=d$ is the physical dimension as well as the number of words.

For $d=2$, there is only one interaction with two levels, whose cooperative relation is simple and trivial. Hence, the qutrit with $d=3$ is the minimal local dimension to consider intriguing cooperative interactions. Throughout this letter, we take three words `\textsf{you}', `\textsf{are}' and `\textsf{here}' as a concrete example to describe the $d=3$ algorithm. It is clear that the corpus is consisted of two allowed sentences `\textsf{you are here}' and inverted `\textsf{here you are}' for these words (questions are not within the scope of this work). Using the string diagram with the pregroup reductions, as shown in Fig.~\ref{fig1}(a), two sentences with grammatical reduction is generated and only one $s$-wire is left open \cite{2007pregroup,2010arxivpregroup,2016arxivpregroup,2023qmipregroup}. A closer correlation between \textsf{you} and \textsf{are} can be found, that is, they tend to stay together and \textsf{here} cannot be inserted into the middle. For $2$-gram, CS is the superposition of all adjacent pairs, producing an entangled state with a normalized coefficient as
\begin{equation}\label{eq1}
	\begin{aligned}
		|{\rm CS}\rangle&=\sum_{\rm pair} |\textsf{word}_1\rangle|\textsf{word}_2\rangle \\
		&=(2|\textsf{you}\rangle|\textsf{are}\rangle+|\textsf{are}\rangle|\textsf{here}\rangle+|\textsf{here}\rangle|\textsf{you}\rangle)/\sqrt{6}.
	\end{aligned}
\end{equation}
It is worth noting that, this combination of CS is similar to the valence bond solid state of Affleck-Kennedy-Lieb-Tasaki (AKLT) model \cite{1987prlaklt,1988cmpaklt,2018prbaklt}. Now the interaction is embodied in the $|{\rm CS}\rangle$ to determine the cooperative evolution. To clarify this step, a more complex example is given in Supplemental Material (SM).

The Frobenius distance is widely used to evaluate the difference between two states \cite{2013prbdistance}. Here, the average distance between $|{\rm CS}\rangle$ and the updating state $\rho$ is defined as
\begin{equation}\label{eq2}
	D=\frac{1}{L}\sum_j \Vert \rho_{j,j+1}-|{\rm CS}\rangle \langle{\rm CS}| \Vert _{F},
\end{equation}
where $\rho_{j,j+1}={\rm tr}_{\overline{j,j+1}}\rho$ is the nearest-neighbor two-site reduced density matrix by tracing out the complement $\overline{j,j+1}$. Eq.~(\ref{eq2}) acts as the postselection indicator to decide whether to update the state. If a random evolution leads to the decreasing of $D$, this update is allowed, otherwise the state before evolution is preserved to next step.

A circuit starts from a trivial product state composed of random vectors in the $d$-dimensional complex space which are denoted as inverted triangles. The random dynamics is performed by a cellular automaton model that replaces the continuous time-evolution under Hamiltonian with a circuit of discrete random local gates \cite{2020quantumca,2021JPAca}, comprising staggered layers of two-site unitary gates acting on even and odd spatial bonds as well as a layer of local one-site gates between them, as schematically shown in Fig.~\ref{fig1}(b). These gates are represented as discrete unitary operators \cite{2023scipostca},
\begin{equation}\label{eq3}
	U_j={\rm exp}\left( -i\frac{\pi}{2}h_j \right),\ U_{j,j+1}={\rm exp}\left( -i\frac{\pi}{2}h_{j,j+1} \right),
\end{equation}
where $h_j$ and $h_{j,j+1}$ are random one-site and two-site operations independent of space or time. Product states undergo a cooperative evolution via these gates. Let us now redefine three-level $X$ and $Z$ matrices different from traditional Pauli matrices of spin-1 to ensure three words are equivalent for every step before judgment of update. Explicitly, the $X$ is given by the sum of the Single-Shift gate and Dual-Shift gate \cite{2007jsaqutrit,2017qipqutrit},
\begin{equation}\label{eq4}
	X= \sum_{\textsf{word}_1\ne \textsf{word}_2}|\textsf{word}_1\rangle\langle\textsf{word}_2|
	=\begin{pmatrix}
		0 & 1 & 1 \\
		1 & 0 & 1 \\
		1 & 1 & 0 \\
	\end{pmatrix}.
\end{equation}
Three $Z$ matrices are expected to contribute identical phases for respective word as
\begin{equation}\label{eq5}
	\begin{aligned}
		Z^{\textsf{you}}&=N^{\textsf{you}}-N^{\textsf{are}}-N^{\textsf{here}}, \\
		Z^{\textsf{are}}&=N^{\textsf{are}}-N^{\textsf{you}}-N^{\textsf{here}}, \\
		Z^{\textsf{here}}&=N^{\textsf{here}}-N^{\textsf{you}}-N^{\textsf{are}},
	\end{aligned}	
\end{equation}
with the occupation $N^{\textsf{word}}= |\textsf{word}\rangle\langle\textsf{word}|$. Analogously, two-site controlled-flip operators are defined as
\begin{equation}\label{eq6}
	\begin{aligned}
		CF^{\textsf{you}}&=N^{\textsf{you}}\otimes X+N^{\textsf{are}}\otimes I+N^{\textsf{here}}\otimes I, \\
		CF^{\textsf{are}}&=N^{\textsf{are}}\otimes X+N^{\textsf{you}}\otimes I+N^{\textsf{here}}\otimes I, \\
		CF^{\textsf{here}}&=N^{\textsf{here}}\otimes X+N^{\textsf{you}}\otimes I+N^{\textsf{are}}\otimes I,
	\end{aligned}	
\end{equation}
with identity matrix $I$ and both two sites can be the target site. Hadamard gate and phase gate which also belong to the Clifford group are not taken into account due to the complex phase. All random operations are randomly sampled from the sets, i.e.
\begin{equation}\label{eq7}
	\begin{aligned}
	h_j&\in\left\{ X, Z^{\textsf{you}},Z^{\textsf{are}} ,Z^{\textsf{here}}, I \right\}, \\
	h_{j,j+1}&\in\left\{ CF^{\textsf{you}}, CF^{\textsf{are}}, CF^{\textsf{here}}, I\otimes I \right\}.
\end{aligned}
\end{equation}
where last identity operators mean no operation and the others are called effective operations. We introduce the gate entropy $S_{\rm os}$($S_{\rm ts}$) to describe the sampling probability $P_{\rm os(ts)}$ of effective one-site (two-site) operations, $S_{\rm os(ts)}=-\log_{10} (P_{\rm os(ts)})$. For instance, $S_{\rm os}=-\log_{10} (1)=0$ indicates that one-site gate $U_j$ is selected with equal probability from operations $\left \{ X, Z^{\textsf{you}}, Z^{\textsf{are}} ,Z^{\textsf{here}} \right \}$ without $I$.

\textit{Results.--} Due to the randomness of initial states and gates sampling, calculation results are averaged over 500 configurations. To reduce finite-size effects of small systems, periodic boundary conditions (PBC) are used for two-site gates and numerical results.

\begin{figure}[htbp]
	\includegraphics[scale=1]{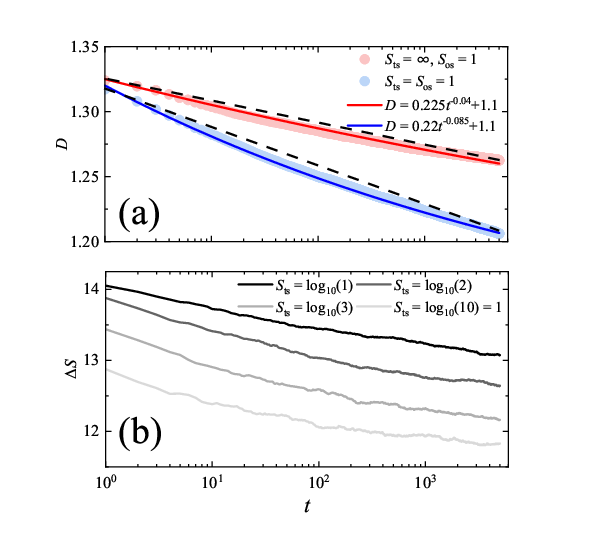}
	\caption{\label{fig2} Update processes with chain $L=6$ in the logarithmic scale, with total steps $t\sim 5\times 10^3$. (a) Evolution of Frobenius distance $D$. Red: $S_{\rm ts}=\infty$ and $S_{\rm os}=1$. Blue: $S_{\rm ts}=S_{\rm os}=1$. Power law fitting results are plotted with solid curves and dashed lines are logarithmic guides for eyes. (b) Evolution of entanglement asymmetry $\Delta S$ with fixed one-site gate $S_{\rm os}=1$ and various two-site gates $S_{\rm ts}=\log_{10} (1), \log_{10} (2), \log_{10} (3)$ and $1$, respectively. }
\end{figure}

A central theme in this model is to understand how the quantum correlation stemming from two-site gates affects the iterative search process and update rate. We now first discuss simplified circuits without effective $U_{j,j+1}$ (equivalent to $S_{\rm ts}=\infty$). This situation can be regarded as a classical evolution under gates $U_j$ which map any product state to another product state without generating entanglement between sites. Roughly speaking, it is similar to the ground state search with the Monte Carlo method in the Ising model \cite{2015paMonteCarlo}. As a straight comparison, we show the evolution of two circuits with the same $L=6$ in Fig.~\ref{fig2}(a). With the aid of dashed lines, we can observe that the the evolution of $D$ is a smooth concave function under the logarithmic auxiliary line, indicating it decreases more rapid than the logarithmic scale. Here, we assume a power law between $D$ and step $t$ as a tentative scenario,
\begin{equation}\label{eq8}
	D=at^{b}+c,
\end{equation}
where $a$, $b$ and $c$ are desired fitting coefficients. The results plotted with solid curves fit well on these decreasing processes. Under $S_{\rm ts}=\infty$, the red solid curve is nearly identical to the dashed line, approximately equal to a logarithmic rate. With the addition of effective two-site gates, the curvature of concave function increases, corresponding to decreasing $b$.
Normally, $a$ is the average initial distance and the constant $c$ independent of gates should be zero, because the distance after infinite steps is supposed to vanish. However, it is impossible to get a $D=0$ state in this discrete cellular automaton and $c=1.1$ might be the allowed minimum. We also fit $D=at^{b}$ to obtain $|b|<0.01$ which is too close to zero and not plotted.

The entanglement asymmetry (EA) serves as a potent tool for evaluating symmetry breaking in many-body systems with globally conserved charges \cite{2022praea,2023ncea,2024quantumea}, which is frequently exploited to study anomalous symmetry restoration in the quantum Mpemba effect \cite{2024prlea}. Despite the absence of conserved charges in these random circuits with $h_j$ and $h_{j,j+1}$, EA remains capable of capturing the evolutionary trends of block distribution of density matrix. We modify the traditional EA with 2-R\'{e}nyi entropy $S(\rho)=-\log[{\rm Tr}(\rho^2)]$ into the sum of three words,
\begin{equation}\label{eq9}
	\begin{aligned}
		\Delta S&=\sum_{j}\sum_{\textsf{word}}\Delta S^{\textsf{word}}_{j,j+1},\\
		\Delta S^{\textsf{word}}_{j,j+1}&=S(\rho_{j,j+1})-S(\rho_{j,j+1;\textsf{word}}),
	\end{aligned}
\end{equation}
where $\rho_{j,j+1;\textsf{word}}=\sum_{q}\Pi^{\textsf{word}}_q \rho_{j,j+1}\Pi^{\textsf{word}}_q$ is the block associated with every symmetry sector of \textsf{word} and $\Pi^{\textsf{word}}_q$ projects $\rho_{j,j+1}$ into the subspace with integer charge $q$. The resulting $\rho_{j,j+1;\textsf{word}}$ are block-diagonal with the eigenbases of two-site state. See SM for the representation of $\Delta S$ in more details. Fig.~\ref{fig2}(b) shows the step dependence of EA with four different two-site gate entropies. One can view that the $\Delta S$ is decreasing following the update of states whose rate is proportional to the number of effective gates, showing the block diagonalization $\rho_{j,j+1}\rightarrow\rho_{j,j+1;\textsf{word}}$.
In other words, the postselection from Eq.~(\ref{eq2}) acts as local kinetical constraints to evolution and squeezes the phase space of many-body systems, making unordered initial states into order related to the CS. Additionally, it can be conjectured that the density matrix is always inclined to be block-diagonal for an appropriate corpus. When given a system with fragmented Hilbert spaces \cite{2020prxfrag,2022prrfrag,2022rppfrag}, the iteration can be used to filter for quantum states in a specified subspace.

\begin{figure}[htbp]
	\includegraphics[scale=1]{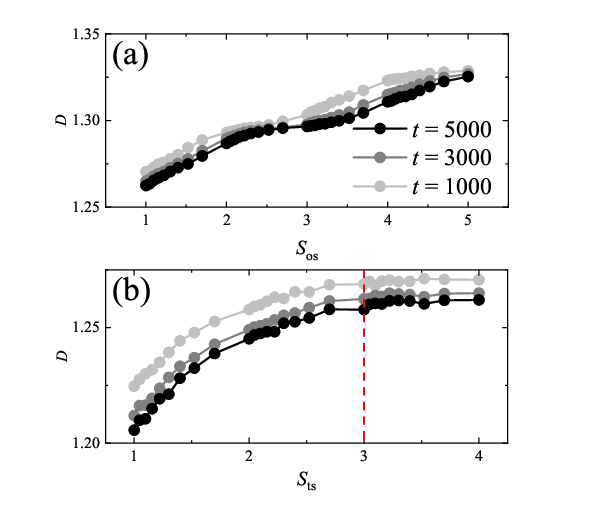}
	\caption{\label{fig3} Final-state average Frobenius distance $D$ at steps $1000$, $3000$, $5000$ with system size $L=6$ as a function of gate entropy (a) $S_{\rm os}$ without effective two-site gate $S_{\rm ts}=\infty$ and (b) $S_{\rm ts}$ with fixed $S_{\rm os}=1$. The dashed line denotes the two-site percolation threshold $S_{\rm ts}=3$.}
\end{figure}

In order to figure out how these gates affect the evolution, we further compare the final states for two circuits. Here what we are concerned about is the update rate and final states do not signify saturation. As seen in Fig.~\ref{fig3}(a), the final-state distance as a function of $S_{\rm os}$ in circuit $S_{\rm ts}=\infty$, shows a rough linear growth. Then too few effective gates ($S_{\rm os}=5$) take three curves back to the same point $D\approx1.33$, the average initial distance to CS. By contrast, for $S_{\rm ts}\neq\infty$ in Fig.~\ref{fig3}(b) the final-state distance before dashed line monotonically increases with $S_{\rm ts}$ in logarithmic from and it is stagnant after the crossover around $S_{\rm ts}=3$, which can be regarded as a two-site bond percolation threshold \cite{1980cmppercolation,2015prpercolation,2021prlpercolation}. Namely, the update for $S_{\rm ts}>3$ is totally determined by one-site gates (see SM for more numerical results).

\begin{figure}[htbp]
	\includegraphics[scale=1]{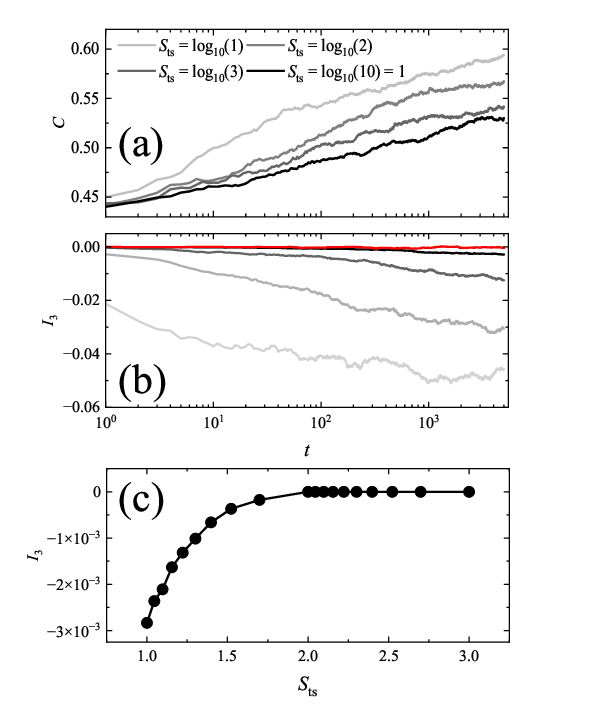}
	\caption{\label{fig4} Evolution of (a) three-site correlation function $C$ and (b) TMI $I_3$ with fixed one-site gate $S_{\rm os}=1$, the system size $L=6$ and total steps $t\sim 5\times 10^3$ in the logarithmic scale. The red curve shows the evolution of TMI with a modified update indicator $D'$ (see main text for definition). (c) TMI of the final state at $t=5000$ as a function of $S_{\rm ts}$. }
\end{figure}

Next, let us consider interesting non-nearest neighbor correlation resulting from local cooperative interactions. As seen in Fig.~\ref{fig4}(a), we plot the three-site correlation function with correct grammar,
\begin{equation}\label{eq10}
	C=\sum_j\langle N^{\textsf{you}}_j N^{\textsf{are}}_{j+1} N^{\textsf{here}}_{j+2}\rangle+\langle N^{\textsf{here}}_j N^{\textsf{you}}_{j+1} N^{\textsf{are}}_{j+2}\rangle.
\end{equation}
and observe monotonous growth. Therefore, more effective sentences composed of three words emerge through the local $2$-gram iteration, verifying the validity of iterative search.

Notably, the cooperative bond between `\textsf{you}' and `\textsf{are}' in Fig.~\ref{fig1}(a) provides a channel allowing the quantum correlation spread. The information from one-site gates is spreading via two-site gates and the bond $|\textsf{you}\rangle|\textsf{are}\rangle$ plays a major role. Furthermore, motivated by the three-word structures, we next consider the delocalized quantum information quantified by the tripartite mutual information (TMI) \cite{2016jheptmi,2016jhepchaos,2018pretmi}, which has been be utilized to characterize long-range entanglement in topological orders \cite{2006prlTopological}. The TMI among three adjacent sites as subsystems is defined by the bipartite mutual information between two sites $I_2(\rho_j:\rho_{k})=S(\rho_j)+S(\rho_{k})-S(\rho_{j,k})$ as
\begin{equation}\label{eq11}
	\begin{aligned}
		I_3(\rho_j:\rho_{j+1}:\rho_{j+2})=&I_2(\rho_j:\rho_{j+1})+I_2(\rho_j:\rho_{j+2})\\
		&-I_2(\rho_j:\rho_{j+1,j+2}),
	\end{aligned}
\end{equation}
and the total TMI is given as $I_3=\sum_j I_3(\rho_j:\rho_{j+1}:\rho_{j+2})$ in PBC. A negative TMI implies that the joint system $(j+1:j+2)$ stores more information about $j$ than the sum of site $j+1$ and $j+2$ containing individually \cite{2018prascrambling,2019prbscrambling,2024prxqscrambling}. As shown in Fig.~\ref{fig4}(b),  TMI decreases gradually and remains negative, which indicates the update scrambles information of any site into others and quantum correlation exists among sites, corresponding to the final state in Fig.~\ref{fig1}(b). In addition, we compute the evolution with four different two-site gate distributions and one can notice that the variation of TMI decays more rapidly with increasing $S_{\rm ts}$ compared to that of the correlation function and EA. In order to further see the influence of cooperative interactions, we modify the indicator Eq.(\ref{eq2}) from summing up each site to grouping every two sites, i.e. $D'=\frac{2}{L}\sum_{k=2j-1}\Vert \rho_{k,k+1}-|{\rm CS}\rangle \langle{\rm CS}| \Vert _{F}$. There then appears a non-interaction bond for any three consecutive sites that the quantum information cannot be transmitted. The resulting TMI with $S_{\rm ts}=1$ plotted in red manifests the scrambling is suppressed. To summarize, the implementation of effective two-site gates directionally optimizes initial states to the target many-body entangled state. It can even be argued that this search process is a kind of information spreading.

To further explore the relation between two-site gates and quantum correlation, we evaluate TMI of the final state as a function of $S_{\rm ts}$ in Fig.~\ref{fig4}(c). For $S_{\rm ts}<2$, the final-state TMI is positively associated with the gate entropy. However, when $S_{\rm ts}>2$, the TMI vanishes and remains at zero, meaning $S_{\rm ts}=2$ is the critical point of information scrambling as well as the threshold of bond percolation. Namely, even though there are interactions in the cellular automaton, the quantum correlation cannot be transmitted to other sites during updating with too few effective interactions ($S_{\rm ts}>2$). It is worth noting that this threshold is not in agreement with the results in Fig.~\ref{fig3}(b), indicating two-site gates still work to update but the information is held and localized in the middle region ($2<S_{\rm ts}<3$). In consequence, the local indicator solely on two sites manifests advantage in the generation of the long-range quantum correlation.

\textit{Perspectives.--} In this work, we have proposed a scheme for iterative search process inspired by the $2$-gram model within the framework of a one-dimensional random quantum cellular automata where database-like CS provides local cooperative interactions. Corresponding dynamics is succinctly captured by the Frobenius distance to CS. We concentrate on the quantum correlation generated by two-site gates and reveal the information scrambling analogous to bond percolation.

Our focus in this study has been on a specific three-word sentence with a small state space and simple CS. Nevertheless, our findings motivate future studies about sentence search in the quantum version with larger physical dimensions and more complicated syntactic structures. For instance, the Matrix Product States framework enables the numerical computation of larger systems \cite{2011aopmps,2021rmpmps}. Moreover, every word in the given corpus can be encoded to a binary vector and construct a full meaning space in quantum states \cite{2021mlstbinary}, which facilitates a range of  effective quantum simulation with two-level systems like the Rydberg atoms array and superconducting qubits \cite{2017natsimulator,2015ncsimulator,2024ncsimulator,2024ncssimulator}. The $2$-gram model can also be extended to $n>2$ with higher precision and a modified $|{\rm CS} \rangle$. We hold the opinion that further studies on the cooperative evolution based on state space will provide a new protocol for directed preparation of entangled states in specific subspaces and develop a potential application for practical quantum computing.

\section*{Acknowledgment}

The authors gratefully acknowledge support from the National Natural Science Foundation of China (Grant No.~12374107). This work is partially supported by High Performance Computing Platform of South China University of Technology.

\bibliography{word_v6.bbl}

\begin{thebibliography}{80}%
\makeatletter
\providecommand \@ifxundefined [1]{%
 \@ifx{#1\undefined}
}%
\providecommand \@ifnum [1]{%
 \ifnum #1\expandafter \@firstoftwo
 \else \expandafter \@secondoftwo
 \fi
}%
\providecommand \@ifx [1]{%
 \ifx #1\expandafter \@firstoftwo
 \else \expandafter \@secondoftwo
 \fi
}%
\providecommand \natexlab [1]{#1}%
\providecommand \enquote  [1]{``#1''}%
\providecommand \bibnamefont  [1]{#1}%
\providecommand \bibfnamefont [1]{#1}%
\providecommand \citenamefont [1]{#1}%
\providecommand \href@noop [0]{\@secondoftwo}%
\providecommand \href [0]{\begingroup \@sanitize@url \@href}%
\providecommand \@href[1]{\@@startlink{#1}\@@href}%
\providecommand \@@href[1]{\endgroup#1\@@endlink}%
\providecommand \@sanitize@url [0]{\catcode `\\12\catcode `\$12\catcode
  `\&12\catcode `\#12\catcode `\^12\catcode `\_12\catcode `\%12\relax}%
\providecommand \@@startlink[1]{}%
\providecommand \@@endlink[0]{}%
\providecommand \url  [0]{\begingroup\@sanitize@url \@url }%
\providecommand \@url [1]{\endgroup\@href {#1}{\urlprefix }}%
\providecommand \urlprefix  [0]{URL }%
\providecommand \Eprint [0]{\href }%
\providecommand \doibase [0]{http://dx.doi.org/}%
\providecommand \selectlanguage [0]{\@gobble}%
\providecommand \bibinfo  [0]{\@secondoftwo}%
\providecommand \bibfield  [0]{\@secondoftwo}%
\providecommand \translation [1]{[#1]}%
\providecommand \BibitemOpen [0]{}%
\providecommand \bibitemStop [0]{}%
\providecommand \bibitemNoStop [0]{.\EOS\space}%
\providecommand \EOS [0]{\spacefactor3000\relax}%
\providecommand \BibitemShut  [1]{\csname bibitem#1\endcsname}%
\let\auto@bib@innerbib\@empty
%</preamble>
\bibitem [{\citenamefont {Gong}\ \emph {et~al.}(2019)\citenamefont {Gong},
  \citenamefont {Chen}, \citenamefont {Zheng}, \citenamefont {Wang},
  \citenamefont {Zha}, \citenamefont {Deng}, \citenamefont {Yan}, \citenamefont
  {Rong}, \citenamefont {Wu}, \citenamefont {Li}, \citenamefont {Chen},
  \citenamefont {Zhao}, \citenamefont {Liang}, \citenamefont {Lin},
  \citenamefont {Xu}, \citenamefont {Guo}, \citenamefont {Sun}, \citenamefont
  {Castellano}, \citenamefont {Wang}, \citenamefont {Peng}, \citenamefont {Lu},
  \citenamefont {Zhu},\ and\ \citenamefont {Pan}}]{2019prlscentanglement}%
  \BibitemOpen
  \bibfield  {author} {\bibinfo {author} {\bibfnamefont {M.}~\bibnamefont
  {Gong}}, \bibinfo {author} {\bibfnamefont {M.-C.}\ \bibnamefont {Chen}},
  \bibinfo {author} {\bibfnamefont {Y.}~\bibnamefont {Zheng}}, \bibinfo
  {author} {\bibfnamefont {S.}~\bibnamefont {Wang}}, \bibinfo {author}
  {\bibfnamefont {C.}~\bibnamefont {Zha}}, \bibinfo {author} {\bibfnamefont
  {H.}~\bibnamefont {Deng}}, \bibinfo {author} {\bibfnamefont {Z.}~\bibnamefont
  {Yan}}, \bibinfo {author} {\bibfnamefont {H.}~\bibnamefont {Rong}}, \bibinfo
  {author} {\bibfnamefont {Y.}~\bibnamefont {Wu}}, \bibinfo {author}
  {\bibfnamefont {S.}~\bibnamefont {Li}}, \bibinfo {author} {\bibfnamefont
  {F.}~\bibnamefont {Chen}}, \bibinfo {author} {\bibfnamefont {Y.}~\bibnamefont
  {Zhao}}, \bibinfo {author} {\bibfnamefont {F.}~\bibnamefont {Liang}},
  \bibinfo {author} {\bibfnamefont {J.}~\bibnamefont {Lin}}, \bibinfo {author}
  {\bibfnamefont {Y.}~\bibnamefont {Xu}}, \bibinfo {author} {\bibfnamefont
  {C.}~\bibnamefont {Guo}}, \bibinfo {author} {\bibfnamefont {L.}~\bibnamefont
  {Sun}}, \bibinfo {author} {\bibfnamefont {A.~D.}\ \bibnamefont {Castellano}},
  \bibinfo {author} {\bibfnamefont {H.}~\bibnamefont {Wang}}, \bibinfo {author}
  {\bibfnamefont {C.}~\bibnamefont {Peng}}, \bibinfo {author} {\bibfnamefont
  {C.-Y.}\ \bibnamefont {Lu}}, \bibinfo {author} {\bibfnamefont
  {X.}~\bibnamefont {Zhu}}, \ and\ \bibinfo {author} {\bibfnamefont {J.-W.}\
  \bibnamefont {Pan}},\ }\href {\doibase 10.1103/PhysRevLett.122.110501}
  {\bibfield  {journal} {\bibinfo  {journal} {Phys. Rev. Lett.}\ }\textbf
  {\bibinfo {volume} {122}},\ \bibinfo {pages} {110501} (\bibinfo {year}
  {2019})}\BibitemShut {NoStop}%
\bibitem [{\citenamefont {Cao}\ \emph {et~al.}(2023)\citenamefont {Cao},
  \citenamefont {Wu}, \citenamefont {Chen}, \citenamefont {Gong}, \citenamefont
  {Wu}, \citenamefont {Ye}, \citenamefont {Zha}, \citenamefont {Qian},
  \citenamefont {Ying}, \citenamefont {Guo}, \citenamefont {Zhu}, \citenamefont
  {Huang}, \citenamefont {Zhao}, \citenamefont {Li}, \citenamefont {Wang},
  \citenamefont {Yu}, \citenamefont {Fan}, \citenamefont {Wu}, \citenamefont
  {Su}, \citenamefont {Deng}, \citenamefont {Rong}, \citenamefont {Li},
  \citenamefont {Zhang}, \citenamefont {Chung}, \citenamefont {Liang},
  \citenamefont {Lin}, \citenamefont {Xu}, \citenamefont {Sun}, \citenamefont
  {Guo}, \citenamefont {Li}, \citenamefont {Huo}, \citenamefont {Peng},
  \citenamefont {Lu}, \citenamefont {Yuan}, \citenamefont {Zhu},\ and\
  \citenamefont {Pan}}]{2023nat51entanglement}%
  \BibitemOpen
  \bibfield  {author} {\bibinfo {author} {\bibfnamefont {S.}~\bibnamefont
  {Cao}}, \bibinfo {author} {\bibfnamefont {B.}~\bibnamefont {Wu}}, \bibinfo
  {author} {\bibfnamefont {F.}~\bibnamefont {Chen}}, \bibinfo {author}
  {\bibfnamefont {M.}~\bibnamefont {Gong}}, \bibinfo {author} {\bibfnamefont
  {Y.}~\bibnamefont {Wu}}, \bibinfo {author} {\bibfnamefont {Y.}~\bibnamefont
  {Ye}}, \bibinfo {author} {\bibfnamefont {C.}~\bibnamefont {Zha}}, \bibinfo
  {author} {\bibfnamefont {H.}~\bibnamefont {Qian}}, \bibinfo {author}
  {\bibfnamefont {C.}~\bibnamefont {Ying}}, \bibinfo {author} {\bibfnamefont
  {S.}~\bibnamefont {Guo}}, \bibinfo {author} {\bibfnamefont {Q.}~\bibnamefont
  {Zhu}}, \bibinfo {author} {\bibfnamefont {H.-L.}\ \bibnamefont {Huang}},
  \bibinfo {author} {\bibfnamefont {Y.}~\bibnamefont {Zhao}}, \bibinfo {author}
  {\bibfnamefont {S.}~\bibnamefont {Li}}, \bibinfo {author} {\bibfnamefont
  {S.}~\bibnamefont {Wang}}, \bibinfo {author} {\bibfnamefont {J.}~\bibnamefont
  {Yu}}, \bibinfo {author} {\bibfnamefont {D.}~\bibnamefont {Fan}}, \bibinfo
  {author} {\bibfnamefont {D.}~\bibnamefont {Wu}}, \bibinfo {author}
  {\bibfnamefont {H.}~\bibnamefont {Su}}, \bibinfo {author} {\bibfnamefont
  {H.}~\bibnamefont {Deng}}, \bibinfo {author} {\bibfnamefont {H.}~\bibnamefont
  {Rong}}, \bibinfo {author} {\bibfnamefont {Y.}~\bibnamefont {Li}}, \bibinfo
  {author} {\bibfnamefont {K.}~\bibnamefont {Zhang}}, \bibinfo {author}
  {\bibfnamefont {T.-H.}\ \bibnamefont {Chung}}, \bibinfo {author}
  {\bibfnamefont {F.}~\bibnamefont {Liang}}, \bibinfo {author} {\bibfnamefont
  {J.}~\bibnamefont {Lin}}, \bibinfo {author} {\bibfnamefont {Y.}~\bibnamefont
  {Xu}}, \bibinfo {author} {\bibfnamefont {L.}~\bibnamefont {Sun}}, \bibinfo
  {author} {\bibfnamefont {C.}~\bibnamefont {Guo}}, \bibinfo {author}
  {\bibfnamefont {N.}~\bibnamefont {Li}}, \bibinfo {author} {\bibfnamefont
  {Y.-H.}\ \bibnamefont {Huo}}, \bibinfo {author} {\bibfnamefont {C.-Z.}\
  \bibnamefont {Peng}}, \bibinfo {author} {\bibfnamefont {C.-Y.}\ \bibnamefont
  {Lu}}, \bibinfo {author} {\bibfnamefont {X.}~\bibnamefont {Yuan}}, \bibinfo
  {author} {\bibfnamefont {X.}~\bibnamefont {Zhu}}, \ and\ \bibinfo {author}
  {\bibfnamefont {J.-W.}\ \bibnamefont {Pan}},\ }\href {\doibase
  10.1038/s41586-023-06195-1} {\bibfield  {journal} {\bibinfo  {journal}
  {Nature}\ }\textbf {\bibinfo {volume} {619}},\ \bibinfo {pages} {738}
  (\bibinfo {year} {2023})}\BibitemShut {NoStop}%
\bibitem [{\citenamefont {Adesso}\ \emph {et~al.}(2016)\citenamefont {Adesso},
  \citenamefont {Bromley},\ and\ \citenamefont
  {Cianciaruso}}]{2016jpacorrelation}%
  \BibitemOpen
  \bibfield  {author} {\bibinfo {author} {\bibfnamefont {G.}~\bibnamefont
  {Adesso}}, \bibinfo {author} {\bibfnamefont {T.~R.}\ \bibnamefont {Bromley}},
  \ and\ \bibinfo {author} {\bibfnamefont {M.}~\bibnamefont {Cianciaruso}},\
  }\href {\doibase 10.1088/1751-8113/49/47/473001} {\bibfield  {journal}
  {\bibinfo  {journal} {Journal of Physics A: Mathematical and Theoretical}\
  }\textbf {\bibinfo {volume} {49}},\ \bibinfo {pages} {473001} (\bibinfo
  {year} {2016})}\BibitemShut {NoStop}%
\bibitem [{\citenamefont {Frérot}\ \emph {et~al.}(2023)\citenamefont
  {Frérot}, \citenamefont {Fadel},\ and\ \citenamefont
  {Lewenstein}}]{2023RPPcorrelation}%
  \BibitemOpen
  \bibfield  {author} {\bibinfo {author} {\bibfnamefont {I.}~\bibnamefont
  {Frérot}}, \bibinfo {author} {\bibfnamefont {M.}~\bibnamefont {Fadel}}, \
  and\ \bibinfo {author} {\bibfnamefont {M.}~\bibnamefont {Lewenstein}},\
  }\href {\doibase 10.1088/1361-6633/acf8d7} {\bibfield  {journal} {\bibinfo
  {journal} {Reports on Progress in Physics}\ }\textbf {\bibinfo {volume}
  {86}},\ \bibinfo {pages} {114001} (\bibinfo {year} {2023})}\BibitemShut
  {NoStop}%
\bibitem [{\citenamefont {Islam}\ \emph {et~al.}(2015)\citenamefont {Islam},
  \citenamefont {Ma}, \citenamefont {Preiss}, \citenamefont {Eric~Tai},
  \citenamefont {Lukin}, \citenamefont {Rispoli},\ and\ \citenamefont
  {Greiner}}]{2015natcorrelation}%
  \BibitemOpen
  \bibfield  {author} {\bibinfo {author} {\bibfnamefont {R.}~\bibnamefont
  {Islam}}, \bibinfo {author} {\bibfnamefont {R.}~\bibnamefont {Ma}}, \bibinfo
  {author} {\bibfnamefont {P.~M.}\ \bibnamefont {Preiss}}, \bibinfo {author}
  {\bibfnamefont {M.}~\bibnamefont {Eric~Tai}}, \bibinfo {author}
  {\bibfnamefont {A.}~\bibnamefont {Lukin}}, \bibinfo {author} {\bibfnamefont
  {M.}~\bibnamefont {Rispoli}}, \ and\ \bibinfo {author} {\bibfnamefont
  {M.}~\bibnamefont {Greiner}},\ }\href {\doibase 10.1038/nature15750}
  {\bibfield  {journal} {\bibinfo  {journal} {Nature}\ }\textbf {\bibinfo
  {volume} {528}},\ \bibinfo {pages} {77} (\bibinfo {year} {2015})}\BibitemShut
  {NoStop}%
\bibitem [{\citenamefont {Bohnet}\ \emph {et~al.}(2016)\citenamefont {Bohnet},
  \citenamefont {Sawyer}, \citenamefont {Britton}, \citenamefont {Wall},
  \citenamefont {Rey}, \citenamefont {Foss-Feig},\ and\ \citenamefont
  {Bollinger}}]{2016scicorrelation}%
  \BibitemOpen
  \bibfield  {author} {\bibinfo {author} {\bibfnamefont {J.~G.}\ \bibnamefont
  {Bohnet}}, \bibinfo {author} {\bibfnamefont {B.~C.}\ \bibnamefont {Sawyer}},
  \bibinfo {author} {\bibfnamefont {J.~W.}\ \bibnamefont {Britton}}, \bibinfo
  {author} {\bibfnamefont {M.~L.}\ \bibnamefont {Wall}}, \bibinfo {author}
  {\bibfnamefont {A.~M.}\ \bibnamefont {Rey}}, \bibinfo {author} {\bibfnamefont
  {M.}~\bibnamefont {Foss-Feig}}, \ and\ \bibinfo {author} {\bibfnamefont
  {J.~J.}\ \bibnamefont {Bollinger}},\ }\href {\doibase
  10.1126/science.aad9958} {\bibfield  {journal} {\bibinfo  {journal}
  {Science}\ }\textbf {\bibinfo {volume} {352}},\ \bibinfo {pages} {1297}
  (\bibinfo {year} {2016})}\BibitemShut {NoStop}%
\bibitem [{\citenamefont {Zhu}\ \emph {et~al.}(2022)\citenamefont {Zhu},
  \citenamefont {Sun}, \citenamefont {Gong}, \citenamefont {Chen},
  \citenamefont {Zhang}, \citenamefont {Wu}, \citenamefont {Ye}, \citenamefont
  {Zha}, \citenamefont {Li}, \citenamefont {Guo}, \citenamefont {Qian},
  \citenamefont {Huang}, \citenamefont {Yu}, \citenamefont {Deng},
  \citenamefont {Rong}, \citenamefont {Lin}, \citenamefont {Xu}, \citenamefont
  {Sun}, \citenamefont {Guo}, \citenamefont {Li}, \citenamefont {Liang},
  \citenamefont {Peng}, \citenamefont {Fan}, \citenamefont {Zhu},\ and\
  \citenamefont {Pan}}]{2022prlscrambling}%
  \BibitemOpen
  \bibfield  {author} {\bibinfo {author} {\bibfnamefont {Q.}~\bibnamefont
  {Zhu}}, \bibinfo {author} {\bibfnamefont {Z.-H.}\ \bibnamefont {Sun}},
  \bibinfo {author} {\bibfnamefont {M.}~\bibnamefont {Gong}}, \bibinfo {author}
  {\bibfnamefont {F.}~\bibnamefont {Chen}}, \bibinfo {author} {\bibfnamefont
  {Y.-R.}\ \bibnamefont {Zhang}}, \bibinfo {author} {\bibfnamefont
  {Y.}~\bibnamefont {Wu}}, \bibinfo {author} {\bibfnamefont {Y.}~\bibnamefont
  {Ye}}, \bibinfo {author} {\bibfnamefont {C.}~\bibnamefont {Zha}}, \bibinfo
  {author} {\bibfnamefont {S.}~\bibnamefont {Li}}, \bibinfo {author}
  {\bibfnamefont {S.}~\bibnamefont {Guo}}, \bibinfo {author} {\bibfnamefont
  {H.}~\bibnamefont {Qian}}, \bibinfo {author} {\bibfnamefont {H.-L.}\
  \bibnamefont {Huang}}, \bibinfo {author} {\bibfnamefont {J.}~\bibnamefont
  {Yu}}, \bibinfo {author} {\bibfnamefont {H.}~\bibnamefont {Deng}}, \bibinfo
  {author} {\bibfnamefont {H.}~\bibnamefont {Rong}}, \bibinfo {author}
  {\bibfnamefont {J.}~\bibnamefont {Lin}}, \bibinfo {author} {\bibfnamefont
  {Y.}~\bibnamefont {Xu}}, \bibinfo {author} {\bibfnamefont {L.}~\bibnamefont
  {Sun}}, \bibinfo {author} {\bibfnamefont {C.}~\bibnamefont {Guo}}, \bibinfo
  {author} {\bibfnamefont {N.}~\bibnamefont {Li}}, \bibinfo {author}
  {\bibfnamefont {F.}~\bibnamefont {Liang}}, \bibinfo {author} {\bibfnamefont
  {C.-Z.}\ \bibnamefont {Peng}}, \bibinfo {author} {\bibfnamefont
  {H.}~\bibnamefont {Fan}}, \bibinfo {author} {\bibfnamefont {X.}~\bibnamefont
  {Zhu}}, \ and\ \bibinfo {author} {\bibfnamefont {J.-W.}\ \bibnamefont
  {Pan}},\ }\href {\doibase 10.1103/PhysRevLett.128.160502} {\bibfield
  {journal} {\bibinfo  {journal} {Phys. Rev. Lett.}\ }\textbf {\bibinfo
  {volume} {128}},\ \bibinfo {pages} {160502} (\bibinfo {year}
  {2022})}\BibitemShut {NoStop}%
\bibitem [{\citenamefont {Yuan}\ \emph {et~al.}(2022)\citenamefont {Yuan},
  \citenamefont {Zhang}, \citenamefont {Wang}, \citenamefont {Duan},\ and\
  \citenamefont {Deng}}]{2022prrscrambling}%
  \BibitemOpen
  \bibfield  {author} {\bibinfo {author} {\bibfnamefont {D.}~\bibnamefont
  {Yuan}}, \bibinfo {author} {\bibfnamefont {S.-Y.}\ \bibnamefont {Zhang}},
  \bibinfo {author} {\bibfnamefont {Y.}~\bibnamefont {Wang}}, \bibinfo {author}
  {\bibfnamefont {L.-M.}\ \bibnamefont {Duan}}, \ and\ \bibinfo {author}
  {\bibfnamefont {D.-L.}\ \bibnamefont {Deng}},\ }\href {\doibase
  10.1103/PhysRevResearch.4.023095} {\bibfield  {journal} {\bibinfo  {journal}
  {Phys. Rev. Res.}\ }\textbf {\bibinfo {volume} {4}},\ \bibinfo {pages}
  {023095} (\bibinfo {year} {2022})}\BibitemShut {NoStop}%
\bibitem [{\citenamefont {Kaufman}\ \emph {et~al.}(2016)\citenamefont
  {Kaufman}, \citenamefont {Tai}, \citenamefont {Lukin}, \citenamefont
  {Rispoli}, \citenamefont {Schittko}, \citenamefont {Preiss},\ and\
  \citenamefont {Greiner}}]{2016scithermalization}%
  \BibitemOpen
  \bibfield  {author} {\bibinfo {author} {\bibfnamefont {A.~M.}\ \bibnamefont
  {Kaufman}}, \bibinfo {author} {\bibfnamefont {M.~E.}\ \bibnamefont {Tai}},
  \bibinfo {author} {\bibfnamefont {A.}~\bibnamefont {Lukin}}, \bibinfo
  {author} {\bibfnamefont {M.}~\bibnamefont {Rispoli}}, \bibinfo {author}
  {\bibfnamefont {R.}~\bibnamefont {Schittko}}, \bibinfo {author}
  {\bibfnamefont {P.~M.}\ \bibnamefont {Preiss}}, \ and\ \bibinfo {author}
  {\bibfnamefont {M.}~\bibnamefont {Greiner}},\ }\href {\doibase
  10.1126/science.aaf6725} {\bibfield  {journal} {\bibinfo  {journal}
  {Science}\ }\textbf {\bibinfo {volume} {353}},\ \bibinfo {pages} {794}
  (\bibinfo {year} {2016})}\BibitemShut {NoStop}%
\bibitem [{\citenamefont {Mathew}\ \emph {et~al.}(2020)\citenamefont {Mathew},
  \citenamefont {Silva}, \citenamefont {Jain}, \citenamefont {Mohan},
  \citenamefont {Adroja}, \citenamefont {Sakai}, \citenamefont {Tomy},
  \citenamefont {Banerjee}, \citenamefont {Goreti}, \citenamefont {N.},
  \citenamefont {Singh},\ and\ \citenamefont
  {Jaiswal-Nagar}}]{2020prrmultientanglement}%
  \BibitemOpen
  \bibfield  {author} {\bibinfo {author} {\bibfnamefont {G.}~\bibnamefont
  {Mathew}}, \bibinfo {author} {\bibfnamefont {S.~L.~L.}\ \bibnamefont
  {Silva}}, \bibinfo {author} {\bibfnamefont {A.}~\bibnamefont {Jain}},
  \bibinfo {author} {\bibfnamefont {A.}~\bibnamefont {Mohan}}, \bibinfo
  {author} {\bibfnamefont {D.~T.}\ \bibnamefont {Adroja}}, \bibinfo {author}
  {\bibfnamefont {V.~G.}\ \bibnamefont {Sakai}}, \bibinfo {author}
  {\bibfnamefont {C.~V.}\ \bibnamefont {Tomy}}, \bibinfo {author}
  {\bibfnamefont {A.}~\bibnamefont {Banerjee}}, \bibinfo {author}
  {\bibfnamefont {R.}~\bibnamefont {Goreti}}, \bibinfo {author} {\bibfnamefont
  {A.~V.}\ \bibnamefont {N.}}, \bibinfo {author} {\bibfnamefont
  {R.}~\bibnamefont {Singh}}, \ and\ \bibinfo {author} {\bibfnamefont
  {D.}~\bibnamefont {Jaiswal-Nagar}},\ }\href {\doibase
  10.1103/PhysRevResearch.2.043329} {\bibfield  {journal} {\bibinfo  {journal}
  {Phys. Rev. Res.}\ }\textbf {\bibinfo {volume} {2}},\ \bibinfo {pages}
  {043329} (\bibinfo {year} {2020})}\BibitemShut {NoStop}%
\bibitem [{\citenamefont {Satzinger}\ \emph {et~al.}(2021)\citenamefont
  {Satzinger}, \citenamefont {Liu}, \citenamefont {Smith}, \citenamefont
  {Knapp}, \citenamefont {Newman}, \citenamefont {Jones}, \citenamefont {Chen},
  \citenamefont {Quintana}, \citenamefont {Mi}, \citenamefont {Dunsworth},
  \citenamefont {Gidney}, \citenamefont {Aleiner}, \citenamefont {Arute},
  \citenamefont {Arya}, \citenamefont {Atalaya}, \citenamefont {Babbush},
  \citenamefont {Bardin}, \citenamefont {Barends}, \citenamefont {Basso},
  \citenamefont {Bengtsson}, \citenamefont {Bilmes}, \citenamefont {Broughton},
  \citenamefont {Buckley}, \citenamefont {Buell}, \citenamefont {Burkett},
  \citenamefont {Bushnell}, \citenamefont {Chiaro}, \citenamefont {Collins},
  \citenamefont {Courtney}, \citenamefont {Demura}, \citenamefont {Derk},
  \citenamefont {Eppens}, \citenamefont {Erickson}, \citenamefont {Faoro},
  \citenamefont {Farhi}, \citenamefont {Fowler}, \citenamefont {Foxen},
  \citenamefont {Giustina}, \citenamefont {Greene}, \citenamefont {Gross},
  \citenamefont {Harrigan}, \citenamefont {Harrington}, \citenamefont {Hilton},
  \citenamefont {Hong}, \citenamefont {Huang}, \citenamefont {Huggins},
  \citenamefont {Ioffe}, \citenamefont {Isakov}, \citenamefont {Jeffrey},
  \citenamefont {Jiang}, \citenamefont {Kafri}, \citenamefont {Kechedzhi},
  \citenamefont {Khattar}, \citenamefont {Kim}, \citenamefont {Klimov},
  \citenamefont {Korotkov}, \citenamefont {Kostritsa}, \citenamefont
  {Landhuis}, \citenamefont {Laptev}, \citenamefont {Locharla}, \citenamefont
  {Lucero}, \citenamefont {Martin}, \citenamefont {McClean}, \citenamefont
  {McEwen}, \citenamefont {Miao}, \citenamefont {Mohseni}, \citenamefont
  {Montazeri}, \citenamefont {Mruczkiewicz}, \citenamefont {Mutus},
  \citenamefont {Naaman}, \citenamefont {Neeley}, \citenamefont {Neill},
  \citenamefont {Niu}, \citenamefont {O’Brien}, \citenamefont {Opremcak},
  \citenamefont {Pató}, \citenamefont {Petukhov}, \citenamefont {Rubin},
  \citenamefont {Sank}, \citenamefont {Shvarts}, \citenamefont {Strain},
  \citenamefont {Szalay}, \citenamefont {Villalonga}, \citenamefont {White},
  \citenamefont {Yao}, \citenamefont {Yeh}, \citenamefont {Yoo}, \citenamefont
  {Zalcman}, \citenamefont {Neven}, \citenamefont {Boixo}, \citenamefont
  {Megrant}, \citenamefont {Chen}, \citenamefont {Kelly}, \citenamefont
  {Smelyanskiy}, \citenamefont {Kitaev}, \citenamefont {Knap}, \citenamefont
  {Pollmann},\ and\ \citenamefont {Roushan}}]{2021scilongentanglement}%
  \BibitemOpen
  \bibfield  {author} {\bibinfo {author} {\bibfnamefont {K.~J.}\ \bibnamefont
  {Satzinger}}, \bibinfo {author} {\bibfnamefont {Y.-J.}\ \bibnamefont {Liu}},
  \bibinfo {author} {\bibfnamefont {A.}~\bibnamefont {Smith}}, \bibinfo
  {author} {\bibfnamefont {C.}~\bibnamefont {Knapp}}, \bibinfo {author}
  {\bibfnamefont {M.}~\bibnamefont {Newman}}, \bibinfo {author} {\bibfnamefont
  {C.}~\bibnamefont {Jones}}, \bibinfo {author} {\bibfnamefont
  {Z.}~\bibnamefont {Chen}}, \bibinfo {author} {\bibfnamefont {C.}~\bibnamefont
  {Quintana}}, \bibinfo {author} {\bibfnamefont {X.}~\bibnamefont {Mi}},
  \bibinfo {author} {\bibfnamefont {A.}~\bibnamefont {Dunsworth}}, \bibinfo
  {author} {\bibfnamefont {C.}~\bibnamefont {Gidney}}, \bibinfo {author}
  {\bibfnamefont {I.}~\bibnamefont {Aleiner}}, \bibinfo {author} {\bibfnamefont
  {F.}~\bibnamefont {Arute}}, \bibinfo {author} {\bibfnamefont
  {K.}~\bibnamefont {Arya}}, \bibinfo {author} {\bibfnamefont {J.}~\bibnamefont
  {Atalaya}}, \bibinfo {author} {\bibfnamefont {R.}~\bibnamefont {Babbush}},
  \bibinfo {author} {\bibfnamefont {J.~C.}\ \bibnamefont {Bardin}}, \bibinfo
  {author} {\bibfnamefont {R.}~\bibnamefont {Barends}}, \bibinfo {author}
  {\bibfnamefont {J.}~\bibnamefont {Basso}}, \bibinfo {author} {\bibfnamefont
  {A.}~\bibnamefont {Bengtsson}}, \bibinfo {author} {\bibfnamefont
  {A.}~\bibnamefont {Bilmes}}, \bibinfo {author} {\bibfnamefont
  {M.}~\bibnamefont {Broughton}}, \bibinfo {author} {\bibfnamefont {B.~B.}\
  \bibnamefont {Buckley}}, \bibinfo {author} {\bibfnamefont {D.~A.}\
  \bibnamefont {Buell}}, \bibinfo {author} {\bibfnamefont {B.}~\bibnamefont
  {Burkett}}, \bibinfo {author} {\bibfnamefont {N.}~\bibnamefont {Bushnell}},
  \bibinfo {author} {\bibfnamefont {B.}~\bibnamefont {Chiaro}}, \bibinfo
  {author} {\bibfnamefont {R.}~\bibnamefont {Collins}}, \bibinfo {author}
  {\bibfnamefont {W.}~\bibnamefont {Courtney}}, \bibinfo {author}
  {\bibfnamefont {S.}~\bibnamefont {Demura}}, \bibinfo {author} {\bibfnamefont
  {A.~R.}\ \bibnamefont {Derk}}, \bibinfo {author} {\bibfnamefont
  {D.}~\bibnamefont {Eppens}}, \bibinfo {author} {\bibfnamefont
  {C.}~\bibnamefont {Erickson}}, \bibinfo {author} {\bibfnamefont
  {L.}~\bibnamefont {Faoro}}, \bibinfo {author} {\bibfnamefont
  {E.}~\bibnamefont {Farhi}}, \bibinfo {author} {\bibfnamefont {A.~G.}\
  \bibnamefont {Fowler}}, \bibinfo {author} {\bibfnamefont {B.}~\bibnamefont
  {Foxen}}, \bibinfo {author} {\bibfnamefont {M.}~\bibnamefont {Giustina}},
  \bibinfo {author} {\bibfnamefont {A.}~\bibnamefont {Greene}}, \bibinfo
  {author} {\bibfnamefont {J.~A.}\ \bibnamefont {Gross}}, \bibinfo {author}
  {\bibfnamefont {M.~P.}\ \bibnamefont {Harrigan}}, \bibinfo {author}
  {\bibfnamefont {S.~D.}\ \bibnamefont {Harrington}}, \bibinfo {author}
  {\bibfnamefont {J.}~\bibnamefont {Hilton}}, \bibinfo {author} {\bibfnamefont
  {S.}~\bibnamefont {Hong}}, \bibinfo {author} {\bibfnamefont {T.}~\bibnamefont
  {Huang}}, \bibinfo {author} {\bibfnamefont {W.~J.}\ \bibnamefont {Huggins}},
  \bibinfo {author} {\bibfnamefont {L.~B.}\ \bibnamefont {Ioffe}}, \bibinfo
  {author} {\bibfnamefont {S.~V.}\ \bibnamefont {Isakov}}, \bibinfo {author}
  {\bibfnamefont {E.}~\bibnamefont {Jeffrey}}, \bibinfo {author} {\bibfnamefont
  {Z.}~\bibnamefont {Jiang}}, \bibinfo {author} {\bibfnamefont
  {D.}~\bibnamefont {Kafri}}, \bibinfo {author} {\bibfnamefont
  {K.}~\bibnamefont {Kechedzhi}}, \bibinfo {author} {\bibfnamefont
  {T.}~\bibnamefont {Khattar}}, \bibinfo {author} {\bibfnamefont
  {S.}~\bibnamefont {Kim}}, \bibinfo {author} {\bibfnamefont {P.~V.}\
  \bibnamefont {Klimov}}, \bibinfo {author} {\bibfnamefont {A.~N.}\
  \bibnamefont {Korotkov}}, \bibinfo {author} {\bibfnamefont {F.}~\bibnamefont
  {Kostritsa}}, \bibinfo {author} {\bibfnamefont {D.}~\bibnamefont {Landhuis}},
  \bibinfo {author} {\bibfnamefont {P.}~\bibnamefont {Laptev}}, \bibinfo
  {author} {\bibfnamefont {A.}~\bibnamefont {Locharla}}, \bibinfo {author}
  {\bibfnamefont {E.}~\bibnamefont {Lucero}}, \bibinfo {author} {\bibfnamefont
  {O.}~\bibnamefont {Martin}}, \bibinfo {author} {\bibfnamefont {J.~R.}\
  \bibnamefont {McClean}}, \bibinfo {author} {\bibfnamefont {M.}~\bibnamefont
  {McEwen}}, \bibinfo {author} {\bibfnamefont {K.~C.}\ \bibnamefont {Miao}},
  \bibinfo {author} {\bibfnamefont {M.}~\bibnamefont {Mohseni}}, \bibinfo
  {author} {\bibfnamefont {S.}~\bibnamefont {Montazeri}}, \bibinfo {author}
  {\bibfnamefont {W.}~\bibnamefont {Mruczkiewicz}}, \bibinfo {author}
  {\bibfnamefont {J.}~\bibnamefont {Mutus}}, \bibinfo {author} {\bibfnamefont
  {O.}~\bibnamefont {Naaman}}, \bibinfo {author} {\bibfnamefont
  {M.}~\bibnamefont {Neeley}}, \bibinfo {author} {\bibfnamefont
  {C.}~\bibnamefont {Neill}}, \bibinfo {author} {\bibfnamefont {M.~Y.}\
  \bibnamefont {Niu}}, \bibinfo {author} {\bibfnamefont {T.~E.}\ \bibnamefont
  {O’Brien}}, \bibinfo {author} {\bibfnamefont {A.}~\bibnamefont {Opremcak}},
  \bibinfo {author} {\bibfnamefont {B.}~\bibnamefont {Pató}}, \bibinfo
  {author} {\bibfnamefont {A.}~\bibnamefont {Petukhov}}, \bibinfo {author}
  {\bibfnamefont {N.~C.}\ \bibnamefont {Rubin}}, \bibinfo {author}
  {\bibfnamefont {D.}~\bibnamefont {Sank}}, \bibinfo {author} {\bibfnamefont
  {V.}~\bibnamefont {Shvarts}}, \bibinfo {author} {\bibfnamefont
  {D.}~\bibnamefont {Strain}}, \bibinfo {author} {\bibfnamefont
  {M.}~\bibnamefont {Szalay}}, \bibinfo {author} {\bibfnamefont
  {B.}~\bibnamefont {Villalonga}}, \bibinfo {author} {\bibfnamefont {T.~C.}\
  \bibnamefont {White}}, \bibinfo {author} {\bibfnamefont {Z.}~\bibnamefont
  {Yao}}, \bibinfo {author} {\bibfnamefont {P.}~\bibnamefont {Yeh}}, \bibinfo
  {author} {\bibfnamefont {J.}~\bibnamefont {Yoo}}, \bibinfo {author}
  {\bibfnamefont {A.}~\bibnamefont {Zalcman}}, \bibinfo {author} {\bibfnamefont
  {H.}~\bibnamefont {Neven}}, \bibinfo {author} {\bibfnamefont
  {S.}~\bibnamefont {Boixo}}, \bibinfo {author} {\bibfnamefont
  {A.}~\bibnamefont {Megrant}}, \bibinfo {author} {\bibfnamefont
  {Y.}~\bibnamefont {Chen}}, \bibinfo {author} {\bibfnamefont {J.}~\bibnamefont
  {Kelly}}, \bibinfo {author} {\bibfnamefont {V.}~\bibnamefont {Smelyanskiy}},
  \bibinfo {author} {\bibfnamefont {A.}~\bibnamefont {Kitaev}}, \bibinfo
  {author} {\bibfnamefont {M.}~\bibnamefont {Knap}}, \bibinfo {author}
  {\bibfnamefont {F.}~\bibnamefont {Pollmann}}, \ and\ \bibinfo {author}
  {\bibfnamefont {P.}~\bibnamefont {Roushan}},\ }\href {\doibase
  10.1126/science.abi8378} {\bibfield  {journal} {\bibinfo  {journal}
  {Science}\ }\textbf {\bibinfo {volume} {374}},\ \bibinfo {pages} {1237}
  (\bibinfo {year} {2021})}\BibitemShut {NoStop}%
\bibitem [{\citenamefont {Hirschberg}\ and\ \citenamefont
  {Manning}(2015)}]{2015scinlp}%
  \BibitemOpen
  \bibfield  {author} {\bibinfo {author} {\bibfnamefont {J.}~\bibnamefont
  {Hirschberg}}\ and\ \bibinfo {author} {\bibfnamefont {C.~D.}\ \bibnamefont
  {Manning}},\ }\href {\doibase 10.1126/science.aaa8685} {\bibfield  {journal}
  {\bibinfo  {journal} {Science}\ }\textbf {\bibinfo {volume} {349}},\ \bibinfo
  {pages} {261} (\bibinfo {year} {2015})}\BibitemShut {NoStop}%
\bibitem [{\citenamefont {Gangal}\ \emph {et~al.}(2023)\citenamefont {Gangal},
  \citenamefont {Shrivastava}, \citenamefont {Hussien}, \citenamefont {Singh},
  \citenamefont {Diwakar}, \citenamefont {Joshi}, \citenamefont {Bisht},\ and\
  \citenamefont {Joshi}}]{2023aipnlp}%
  \BibitemOpen
  \bibfield  {author} {\bibinfo {author} {\bibfnamefont {A.}~\bibnamefont
  {Gangal}}, \bibinfo {author} {\bibfnamefont {A.}~\bibnamefont {Shrivastava}},
  \bibinfo {author} {\bibfnamefont {N.~M.}\ \bibnamefont {Hussien}}, \bibinfo
  {author} {\bibfnamefont {P.}~\bibnamefont {Singh}}, \bibinfo {author}
  {\bibfnamefont {M.}~\bibnamefont {Diwakar}}, \bibinfo {author} {\bibfnamefont
  {K.}~\bibnamefont {Joshi}}, \bibinfo {author} {\bibfnamefont
  {S.}~\bibnamefont {Bisht}}, \ and\ \bibinfo {author} {\bibfnamefont {N.~C.}\
  \bibnamefont {Joshi}},\ }\href {\doibase 10.1063/5.0153960} {\bibfield
  {journal} {\bibinfo  {journal} {AIP Conference Proceedings}\ }\textbf
  {\bibinfo {volume} {2771}},\ \bibinfo {pages} {020010} (\bibinfo {year}
  {2023})},\ \Eprint
  {http://arxiv.org/abs/https://pubs.aip.org/aip/acp/article-pdf/doi/10.1063/5.0153960/18105894/020010\_1\_5.0153960.pdf}
  {https://pubs.aip.org/aip/acp/article-pdf/doi/10.1063/5.0153960/18105894/020010\_1\_5.0153960.pdf}
  \BibitemShut {NoStop}%
\bibitem [{\citenamefont {Guarasci}\ \emph {et~al.}(2022)\citenamefont
  {Guarasci}, \citenamefont {De~Pietro},\ and\ \citenamefont
  {Esposito}}]{2022asqnlp}%
  \BibitemOpen
  \bibfield  {author} {\bibinfo {author} {\bibfnamefont {R.}~\bibnamefont
  {Guarasci}}, \bibinfo {author} {\bibfnamefont {G.}~\bibnamefont {De~Pietro}},
  \ and\ \bibinfo {author} {\bibfnamefont {M.}~\bibnamefont {Esposito}},\
  }\href {\doibase 10.3390/app12115651} {\bibfield  {journal} {\bibinfo
  {journal} {Applied Sciences}\ }\textbf {\bibinfo {volume} {12}} (\bibinfo
  {year} {2022}),\ 10.3390/app12115651}\BibitemShut {NoStop}%
\bibitem [{\citenamefont {Zhang}\ \emph {et~al.}(2018)\citenamefont {Zhang},
  \citenamefont {Su}, \citenamefont {Zhang}, \citenamefont {Wang},\ and\
  \citenamefont {Song}}]{2018qnlp}%
  \BibitemOpen
  \bibfield  {author} {\bibinfo {author} {\bibfnamefont {P.}~\bibnamefont
  {Zhang}}, \bibinfo {author} {\bibfnamefont {Z.}~\bibnamefont {Su}}, \bibinfo
  {author} {\bibfnamefont {L.}~\bibnamefont {Zhang}}, \bibinfo {author}
  {\bibfnamefont {B.}~\bibnamefont {Wang}}, \ and\ \bibinfo {author}
  {\bibfnamefont {D.}~\bibnamefont {Song}},\ }in\ \href {\doibase
  10.1145/3269206.3271723} {\emph {\bibinfo {booktitle} {Proceedings of the
  27th ACM International Conference on Information and Knowledge
  Management}}},\ \bibinfo {series and number} {CIKM '18}\ (\bibinfo
  {publisher} {Association for Computing Machinery},\ \bibinfo {address} {New
  York, NY, USA},\ \bibinfo {year} {2018})\ p.\ \bibinfo {pages}
  {1303–1312}\BibitemShut {NoStop}%
\bibitem [{\citenamefont {Chen}\ \emph {et~al.}(2023)\citenamefont {Chen},
  \citenamefont {Pan},\ and\ \citenamefont {Dong}}]{2023IEEEnlp}%
  \BibitemOpen
  \bibfield  {author} {\bibinfo {author} {\bibfnamefont {Y.}~\bibnamefont
  {Chen}}, \bibinfo {author} {\bibfnamefont {Y.}~\bibnamefont {Pan}}, \ and\
  \bibinfo {author} {\bibfnamefont {D.}~\bibnamefont {Dong}},\ }\href {\doibase
  10.1109/TCYB.2021.3131252} {\bibfield  {journal} {\bibinfo  {journal} {IEEE
  Transactions on Cybernetics}\ }\textbf {\bibinfo {volume} {53}},\ \bibinfo
  {pages} {3467} (\bibinfo {year} {2023})}\BibitemShut {NoStop}%
\bibitem [{\citenamefont {Hibat-Allah}\ \emph {et~al.}(2020)\citenamefont
  {Hibat-Allah}, \citenamefont {Ganahl}, \citenamefont {Hayward}, \citenamefont
  {Melko},\ and\ \citenamefont {Carrasquilla}}]{2020prrnw}%
  \BibitemOpen
  \bibfield  {author} {\bibinfo {author} {\bibfnamefont {M.}~\bibnamefont
  {Hibat-Allah}}, \bibinfo {author} {\bibfnamefont {M.}~\bibnamefont {Ganahl}},
  \bibinfo {author} {\bibfnamefont {L.~E.}\ \bibnamefont {Hayward}}, \bibinfo
  {author} {\bibfnamefont {R.~G.}\ \bibnamefont {Melko}}, \ and\ \bibinfo
  {author} {\bibfnamefont {J.}~\bibnamefont {Carrasquilla}},\ }\href {\doibase
  10.1103/PhysRevResearch.2.023358} {\bibfield  {journal} {\bibinfo  {journal}
  {Phys. Rev. Res.}\ }\textbf {\bibinfo {volume} {2}},\ \bibinfo {pages}
  {023358} (\bibinfo {year} {2020})}\BibitemShut {NoStop}%
\bibitem [{\citenamefont {Zhang}\ and\ \citenamefont
  {Di~Ventra}(2023)}]{2023prbTransformer}%
  \BibitemOpen
  \bibfield  {author} {\bibinfo {author} {\bibfnamefont {Y.-H.}\ \bibnamefont
  {Zhang}}\ and\ \bibinfo {author} {\bibfnamefont {M.}~\bibnamefont
  {Di~Ventra}},\ }\href {\doibase 10.1103/PhysRevB.107.075147} {\bibfield
  {journal} {\bibinfo  {journal} {Phys. Rev. B}\ }\textbf {\bibinfo {volume}
  {107}},\ \bibinfo {pages} {075147} (\bibinfo {year} {2023})}\BibitemShut
  {NoStop}%
\bibitem [{\citenamefont {Viteritti}\ \emph {et~al.}(2023)\citenamefont
  {Viteritti}, \citenamefont {Rende},\ and\ \citenamefont
  {Becca}}]{2023prlTransformer}%
  \BibitemOpen
  \bibfield  {author} {\bibinfo {author} {\bibfnamefont {L.~L.}\ \bibnamefont
  {Viteritti}}, \bibinfo {author} {\bibfnamefont {R.}~\bibnamefont {Rende}}, \
  and\ \bibinfo {author} {\bibfnamefont {F.}~\bibnamefont {Becca}},\ }\href
  {\doibase 10.1103/PhysRevLett.130.236401} {\bibfield  {journal} {\bibinfo
  {journal} {Phys. Rev. Lett.}\ }\textbf {\bibinfo {volume} {130}},\ \bibinfo
  {pages} {236401} (\bibinfo {year} {2023})}\BibitemShut {NoStop}%
\bibitem [{\citenamefont {Stephens}\ and\ \citenamefont
  {Bialek}(2010)}]{2010presml}%
  \BibitemOpen
  \bibfield  {author} {\bibinfo {author} {\bibfnamefont {G.~J.}\ \bibnamefont
  {Stephens}}\ and\ \bibinfo {author} {\bibfnamefont {W.}~\bibnamefont
  {Bialek}},\ }\href {\doibase 10.1103/PhysRevE.81.066119} {\bibfield
  {journal} {\bibinfo  {journal} {Phys. Rev. E}\ }\textbf {\bibinfo {volume}
  {81}},\ \bibinfo {pages} {066119} (\bibinfo {year} {2010})}\BibitemShut
  {NoStop}%
\bibitem [{\citenamefont {Corral}\ and\ \citenamefont {García~del
  Muro}(2020)}]{2020entropysml}%
  \BibitemOpen
  \bibfield  {author} {\bibinfo {author} {\bibfnamefont {l.}~\bibnamefont
  {Corral}}\ and\ \bibinfo {author} {\bibfnamefont {M.}~\bibnamefont
  {García~del Muro}},\ }\href {\doibase 10.3390/e22020179} {\bibfield
  {journal} {\bibinfo  {journal} {Entropy}\ }\textbf {\bibinfo {volume} {22}}
  (\bibinfo {year} {2020}),\ 10.3390/e22020179}\BibitemShut {NoStop}%
\bibitem [{\citenamefont {Jurafsky}\ and\ \citenamefont
  {Martin}(2024)}]{2024bookn-gram}%
  \BibitemOpen
  \bibfield  {author} {\bibinfo {author} {\bibfnamefont {D.}~\bibnamefont
  {Jurafsky}}\ and\ \bibinfo {author} {\bibfnamefont {J.~H.}\ \bibnamefont
  {Martin}},\ }\href {https://web.stanford.edu/~jurafsky/slp3/} {\emph
  {\bibinfo {title} {Speech and Language Processing: An Introduction to Natural
  Language Processing, Computational Linguistics, and Speech Recognition with
  Language Models}}},\ \bibinfo {edition} {3rd}\ ed.\ (\bibinfo {year} {2024})\
  \bibinfo {note} {online manuscript released August 20, 2024}\BibitemShut
  {NoStop}%
\bibitem [{\citenamefont {Fink}(2014)}]{2014bookn-gram}%
  \BibitemOpen
  \bibfield  {author} {\bibinfo {author} {\bibfnamefont {G.~A.}\ \bibnamefont
  {Fink}},\ }\href@noop {} {\emph {\bibinfo {title} {Markov models for pattern
  recognition: from theory to applications}}}\ (\bibinfo  {publisher} {Springer
  Science \& Business Media},\ \bibinfo {year} {2014})\BibitemShut {NoStop}%
\bibitem [{\citenamefont {He}\ \emph {et~al.}(2019)\citenamefont {He},
  \citenamefont {Huang}, \citenamefont {Bai},\ and\ \citenamefont
  {Bai}}]{2019IEEEn-gram}%
  \BibitemOpen
  \bibfield  {author} {\bibinfo {author} {\bibfnamefont {X.}~\bibnamefont
  {He}}, \bibinfo {author} {\bibfnamefont {T.}~\bibnamefont {Huang}}, \bibinfo
  {author} {\bibfnamefont {S.}~\bibnamefont {Bai}}, \ and\ \bibinfo {author}
  {\bibfnamefont {X.}~\bibnamefont {Bai}},\ }in\ \href {\doibase
  10.1109/ICCV.2019.00761} {\emph {\bibinfo {booktitle} {2019 IEEE/CVF
  International Conference on Computer Vision (ICCV)}}}\ (\bibinfo {year}
  {2019})\ pp.\ \bibinfo {pages} {7514--7523}\BibitemShut {NoStop}%
\bibitem [{\citenamefont {Lopez-Gazpio}\ \emph {et~al.}(2019)\citenamefont
  {Lopez-Gazpio}, \citenamefont {Maritxalar}, \citenamefont {Lapata},\ and\
  \citenamefont {Agirre}}]{2019esan-gram}%
  \BibitemOpen
  \bibfield  {author} {\bibinfo {author} {\bibfnamefont {I.}~\bibnamefont
  {Lopez-Gazpio}}, \bibinfo {author} {\bibfnamefont {M.}~\bibnamefont
  {Maritxalar}}, \bibinfo {author} {\bibfnamefont {M.}~\bibnamefont {Lapata}},
  \ and\ \bibinfo {author} {\bibfnamefont {E.}~\bibnamefont {Agirre}},\ }\href
  {\doibase https://doi.org/10.1016/j.eswa.2019.04.054} {\bibfield  {journal}
  {\bibinfo  {journal} {Expert Systems with Applications}\ }\textbf {\bibinfo
  {volume} {132}},\ \bibinfo {pages} {1} (\bibinfo {year} {2019})}\BibitemShut
  {NoStop}%
\bibitem [{\citenamefont {Grover}(1996)}]{1996Grover}%
  \BibitemOpen
  \bibfield  {author} {\bibinfo {author} {\bibfnamefont {L.~K.}\ \bibnamefont
  {Grover}},\ }in\ \href {\doibase 10.1145/237814.237866} {\emph {\bibinfo
  {booktitle} {Proceedings of the Twenty-Eighth Annual ACM Symposium on Theory
  of Computing}}},\ \bibinfo {series and number} {STOC '96}\ (\bibinfo
  {publisher} {Association for Computing Machinery},\ \bibinfo {address} {New
  York, NY, USA},\ \bibinfo {year} {1996})\ p.\ \bibinfo {pages}
  {212–219}\BibitemShut {NoStop}%
\bibitem [{\citenamefont {Chuang}\ \emph {et~al.}(1998)\citenamefont {Chuang},
  \citenamefont {Gershenfeld},\ and\ \citenamefont {Kubinec}}]{1998prlsearch}%
  \BibitemOpen
  \bibfield  {author} {\bibinfo {author} {\bibfnamefont {I.~L.}\ \bibnamefont
  {Chuang}}, \bibinfo {author} {\bibfnamefont {N.}~\bibnamefont {Gershenfeld}},
  \ and\ \bibinfo {author} {\bibfnamefont {M.}~\bibnamefont {Kubinec}},\ }\href
  {\doibase 10.1103/PhysRevLett.80.3408} {\bibfield  {journal} {\bibinfo
  {journal} {Phys. Rev. Lett.}\ }\textbf {\bibinfo {volume} {80}},\ \bibinfo
  {pages} {3408} (\bibinfo {year} {1998})}\BibitemShut {NoStop}%
\bibitem [{\citenamefont {Giri}\ and\ \citenamefont
  {Korepin}(2017)}]{2017qifsearch}%
  \BibitemOpen
  \bibfield  {author} {\bibinfo {author} {\bibfnamefont {P.~R.}\ \bibnamefont
  {Giri}}\ and\ \bibinfo {author} {\bibfnamefont {V.~E.}\ \bibnamefont
  {Korepin}},\ }\href {\doibase 10.1007/s11128-017-1768-7} {\bibfield
  {journal} {\bibinfo  {journal} {Quantum Information Processing}\ }\textbf
  {\bibinfo {volume} {16}},\ \bibinfo {pages} {315} (\bibinfo {year}
  {2017})}\BibitemShut {NoStop}%
\bibitem [{\citenamefont {Mandviwalla}\ \emph {et~al.}(2018)\citenamefont
  {Mandviwalla}, \citenamefont {Ohshiro},\ and\ \citenamefont
  {Ji}}]{2018IEEEsearch}%
  \BibitemOpen
  \bibfield  {author} {\bibinfo {author} {\bibfnamefont {A.}~\bibnamefont
  {Mandviwalla}}, \bibinfo {author} {\bibfnamefont {K.}~\bibnamefont
  {Ohshiro}}, \ and\ \bibinfo {author} {\bibfnamefont {B.}~\bibnamefont {Ji}},\
  }in\ \href {\doibase 10.1109/BigData.2018.8622457} {\emph {\bibinfo
  {booktitle} {2018 IEEE International Conference on Big Data (Big Data)}}}\
  (\bibinfo {year} {2018})\ pp.\ \bibinfo {pages} {2531--2537}\BibitemShut
  {NoStop}%
\bibitem [{\citenamefont {Wang}\ and\ \citenamefont
  {Krstic}(2020)}]{2020prasearch}%
  \BibitemOpen
  \bibfield  {author} {\bibinfo {author} {\bibfnamefont {Y.}~\bibnamefont
  {Wang}}\ and\ \bibinfo {author} {\bibfnamefont {P.~S.}\ \bibnamefont
  {Krstic}},\ }\href {\doibase 10.1103/PhysRevA.102.042609} {\bibfield
  {journal} {\bibinfo  {journal} {Phys. Rev. A}\ }\textbf {\bibinfo {volume}
  {102}},\ \bibinfo {pages} {042609} (\bibinfo {year} {2020})}\BibitemShut
  {NoStop}%
\bibitem [{\citenamefont {Niedermeier}\ \emph {et~al.}(2024)\citenamefont
  {Niedermeier}, \citenamefont {Lado},\ and\ \citenamefont
  {Flindt}}]{2024prrsearch}%
  \BibitemOpen
  \bibfield  {author} {\bibinfo {author} {\bibfnamefont {M.}~\bibnamefont
  {Niedermeier}}, \bibinfo {author} {\bibfnamefont {J.~L.}\ \bibnamefont
  {Lado}}, \ and\ \bibinfo {author} {\bibfnamefont {C.}~\bibnamefont
  {Flindt}},\ }\href {\doibase 10.1103/PhysRevResearch.6.033325} {\bibfield
  {journal} {\bibinfo  {journal} {Phys. Rev. Res.}\ }\textbf {\bibinfo {volume}
  {6}},\ \bibinfo {pages} {033325} (\bibinfo {year} {2024})}\BibitemShut
  {NoStop}%
\bibitem [{\citenamefont {Cui}\ and\ \citenamefont
  {Fan}(2010)}]{2010jpasearch}%
  \BibitemOpen
  \bibfield  {author} {\bibinfo {author} {\bibfnamefont {J.}~\bibnamefont
  {Cui}}\ and\ \bibinfo {author} {\bibfnamefont {H.}~\bibnamefont {Fan}},\
  }\href {\doibase 10.1088/1751-8113/43/4/045305} {\bibfield  {journal}
  {\bibinfo  {journal} {Journal of Physics A: Mathematical and Theoretical}\
  }\textbf {\bibinfo {volume} {43}},\ \bibinfo {pages} {045305} (\bibinfo
  {year} {2010})}\BibitemShut {NoStop}%
\bibitem [{\citenamefont {Dunn}\ \emph {et~al.}(2020)\citenamefont {Dunn},
  \citenamefont {Wang}, \citenamefont {Ganose}, \citenamefont {Dopp},\ and\
  \citenamefont {Jain}}]{2020npjsearch}%
  \BibitemOpen
  \bibfield  {author} {\bibinfo {author} {\bibfnamefont {A.}~\bibnamefont
  {Dunn}}, \bibinfo {author} {\bibfnamefont {Q.}~\bibnamefont {Wang}}, \bibinfo
  {author} {\bibfnamefont {A.}~\bibnamefont {Ganose}}, \bibinfo {author}
  {\bibfnamefont {D.}~\bibnamefont {Dopp}}, \ and\ \bibinfo {author}
  {\bibfnamefont {A.}~\bibnamefont {Jain}},\ }\href {\doibase
  10.1038/s41524-020-00406-3} {\bibfield  {journal} {\bibinfo  {journal} {npj
  Computational Materials}\ }\textbf {\bibinfo {volume} {6}},\ \bibinfo {pages}
  {138} (\bibinfo {year} {2020})}\BibitemShut {NoStop}%
\bibitem [{\citenamefont {Li}\ \emph {et~al.}(2023)\citenamefont {Li},
  \citenamefont {Liang}, \citenamefont {Zhang}, \citenamefont {Lin},\ and\
  \citenamefont {Wei}}]{2023npjsearch}%
  \BibitemOpen
  \bibfield  {author} {\bibinfo {author} {\bibfnamefont {C.-N.}\ \bibnamefont
  {Li}}, \bibinfo {author} {\bibfnamefont {H.-P.}\ \bibnamefont {Liang}},
  \bibinfo {author} {\bibfnamefont {X.}~\bibnamefont {Zhang}}, \bibinfo
  {author} {\bibfnamefont {Z.}~\bibnamefont {Lin}}, \ and\ \bibinfo {author}
  {\bibfnamefont {S.-H.}\ \bibnamefont {Wei}},\ }\href {\doibase
  10.1038/s41524-023-01122-4} {\bibfield  {journal} {\bibinfo  {journal} {npj
  Computational Materials}\ }\textbf {\bibinfo {volume} {9}},\ \bibinfo {pages}
  {176} (\bibinfo {year} {2023})}\BibitemShut {NoStop}%
\bibitem [{\citenamefont {Wang}\ \emph {et~al.}(2023)\citenamefont {Wang},
  \citenamefont {Gao}, \citenamefont {Han}, \citenamefont {Ding}, \citenamefont
  {Pan}, \citenamefont {Wang}, \citenamefont {Jia}, \citenamefont {Wang},
  \citenamefont {Xing},\ and\ \citenamefont {Sun}}]{2023nsrsearch}%
  \BibitemOpen
  \bibfield  {author} {\bibinfo {author} {\bibfnamefont {J.}~\bibnamefont
  {Wang}}, \bibinfo {author} {\bibfnamefont {H.}~\bibnamefont {Gao}}, \bibinfo
  {author} {\bibfnamefont {Y.}~\bibnamefont {Han}}, \bibinfo {author}
  {\bibfnamefont {C.}~\bibnamefont {Ding}}, \bibinfo {author} {\bibfnamefont
  {S.}~\bibnamefont {Pan}}, \bibinfo {author} {\bibfnamefont {Y.}~\bibnamefont
  {Wang}}, \bibinfo {author} {\bibfnamefont {Q.}~\bibnamefont {Jia}}, \bibinfo
  {author} {\bibfnamefont {H.-T.}\ \bibnamefont {Wang}}, \bibinfo {author}
  {\bibfnamefont {D.}~\bibnamefont {Xing}}, \ and\ \bibinfo {author}
  {\bibfnamefont {J.}~\bibnamefont {Sun}},\ }\href {\doibase
  10.1093/nsr/nwad128} {\bibfield  {journal} {\bibinfo  {journal} {National
  Science Review}\ }\textbf {\bibinfo {volume} {10}},\ \bibinfo {pages}
  {nwad128} (\bibinfo {year} {2023})},\ \Eprint
  {http://arxiv.org/abs/https://academic.oup.com/nsr/article-pdf/10/7/nwad128/50709989/nwad128.pdf}
  {https://academic.oup.com/nsr/article-pdf/10/7/nwad128/50709989/nwad128.pdf}
  \BibitemShut {NoStop}%
\bibitem [{\citenamefont {Ci}\ \emph {et~al.}(2010)\citenamefont {Ci},
  \citenamefont {Song}, \citenamefont {Jin}, \citenamefont {Jariwala},
  \citenamefont {Wu}, \citenamefont {Li}, \citenamefont {Srivastava},
  \citenamefont {Wang}, \citenamefont {Storr}, \citenamefont {Balicas},
  \citenamefont {Liu},\ and\ \citenamefont {Ajayan}}]{2010nmbcn}%
  \BibitemOpen
  \bibfield  {author} {\bibinfo {author} {\bibfnamefont {L.}~\bibnamefont
  {Ci}}, \bibinfo {author} {\bibfnamefont {L.}~\bibnamefont {Song}}, \bibinfo
  {author} {\bibfnamefont {C.}~\bibnamefont {Jin}}, \bibinfo {author}
  {\bibfnamefont {D.}~\bibnamefont {Jariwala}}, \bibinfo {author}
  {\bibfnamefont {D.}~\bibnamefont {Wu}}, \bibinfo {author} {\bibfnamefont
  {Y.}~\bibnamefont {Li}}, \bibinfo {author} {\bibfnamefont {A.}~\bibnamefont
  {Srivastava}}, \bibinfo {author} {\bibfnamefont {Z.~F.}\ \bibnamefont
  {Wang}}, \bibinfo {author} {\bibfnamefont {K.}~\bibnamefont {Storr}},
  \bibinfo {author} {\bibfnamefont {L.}~\bibnamefont {Balicas}}, \bibinfo
  {author} {\bibfnamefont {F.}~\bibnamefont {Liu}}, \ and\ \bibinfo {author}
  {\bibfnamefont {P.~M.}\ \bibnamefont {Ajayan}},\ }\href {\doibase
  10.1038/nmat2711} {\bibfield  {journal} {\bibinfo  {journal} {Nature
  Materials}\ }\textbf {\bibinfo {volume} {9}},\ \bibinfo {pages} {430}
  (\bibinfo {year} {2010})}\BibitemShut {NoStop}%
\bibitem [{\citenamefont {Lu}\ \emph {et~al.}(2013)\citenamefont {Lu},
  \citenamefont {Zhang}, \citenamefont {Feng~Liu}, \citenamefont {Zhang},
  \citenamefont {Chien~Sum}, \citenamefont {Castro~Neto},\ and\ \citenamefont
  {Loh}}]{2013ncbcn}%
  \BibitemOpen
  \bibfield  {author} {\bibinfo {author} {\bibfnamefont {J.}~\bibnamefont
  {Lu}}, \bibinfo {author} {\bibfnamefont {K.}~\bibnamefont {Zhang}}, \bibinfo
  {author} {\bibfnamefont {X.}~\bibnamefont {Feng~Liu}}, \bibinfo {author}
  {\bibfnamefont {H.}~\bibnamefont {Zhang}}, \bibinfo {author} {\bibfnamefont
  {T.}~\bibnamefont {Chien~Sum}}, \bibinfo {author} {\bibfnamefont {A.~H.}\
  \bibnamefont {Castro~Neto}}, \ and\ \bibinfo {author} {\bibfnamefont {K.~P.}\
  \bibnamefont {Loh}},\ }\href {\doibase 10.1038/ncomms3681} {\bibfield
  {journal} {\bibinfo  {journal} {Nature Communications}\ }\textbf {\bibinfo
  {volume} {4}},\ \bibinfo {pages} {2681} (\bibinfo {year} {2013})}\BibitemShut
  {NoStop}%
\bibitem [{\citenamefont {Liu}\ \emph {et~al.}(2016)\citenamefont {Liu},
  \citenamefont {Zhong}, \citenamefont {Yan},\ and\ \citenamefont
  {Wang}}]{2016pccpbcn}%
  \BibitemOpen
  \bibfield  {author} {\bibinfo {author} {\bibfnamefont {Z.}~\bibnamefont
  {Liu}}, \bibinfo {author} {\bibfnamefont {X.}~\bibnamefont {Zhong}}, \bibinfo
  {author} {\bibfnamefont {H.}~\bibnamefont {Yan}}, \ and\ \bibinfo {author}
  {\bibfnamefont {R.-Z.}\ \bibnamefont {Wang}},\ }\href {\doibase
  10.1039/C5CP06037K} {\bibfield  {journal} {\bibinfo  {journal} {Phys. Chem.
  Chem. Phys.}\ }\textbf {\bibinfo {volume} {18}},\ \bibinfo {pages} {974}
  (\bibinfo {year} {2016})}\BibitemShut {NoStop}%
\bibitem [{\citenamefont {Meakin}\ \emph {et~al.}(1986)\citenamefont {Meakin},
  \citenamefont {Ramanlal}, \citenamefont {Sander},\ and\ \citenamefont
  {Ball}}]{1986prakpz}%
  \BibitemOpen
  \bibfield  {author} {\bibinfo {author} {\bibfnamefont {P.}~\bibnamefont
  {Meakin}}, \bibinfo {author} {\bibfnamefont {P.}~\bibnamefont {Ramanlal}},
  \bibinfo {author} {\bibfnamefont {L.~M.}\ \bibnamefont {Sander}}, \ and\
  \bibinfo {author} {\bibfnamefont {R.~C.}\ \bibnamefont {Ball}},\ }\href
  {\doibase 10.1103/PhysRevA.34.5091} {\bibfield  {journal} {\bibinfo
  {journal} {Phys. Rev. A}\ }\textbf {\bibinfo {volume} {34}},\ \bibinfo
  {pages} {5091} (\bibinfo {year} {1986})}\BibitemShut {NoStop}%
\bibitem [{\citenamefont {Tang}\ and\ \citenamefont
  {Liang}(1993)}]{1993prlkpz}%
  \BibitemOpen
  \bibfield  {author} {\bibinfo {author} {\bibfnamefont {C.}~\bibnamefont
  {Tang}}\ and\ \bibinfo {author} {\bibfnamefont {S.}~\bibnamefont {Liang}},\
  }\href {\doibase 10.1103/PhysRevLett.71.2769} {\bibfield  {journal} {\bibinfo
   {journal} {Phys. Rev. Lett.}\ }\textbf {\bibinfo {volume} {71}},\ \bibinfo
  {pages} {2769} (\bibinfo {year} {1993})}\BibitemShut {NoStop}%
\bibitem [{\citenamefont {Nahum}\ \emph {et~al.}(2017)\citenamefont {Nahum},
  \citenamefont {Ruhman}, \citenamefont {Vijay},\ and\ \citenamefont
  {Haah}}]{2017prxkpz}%
  \BibitemOpen
  \bibfield  {author} {\bibinfo {author} {\bibfnamefont {A.}~\bibnamefont
  {Nahum}}, \bibinfo {author} {\bibfnamefont {J.}~\bibnamefont {Ruhman}},
  \bibinfo {author} {\bibfnamefont {S.}~\bibnamefont {Vijay}}, \ and\ \bibinfo
  {author} {\bibfnamefont {J.}~\bibnamefont {Haah}},\ }\href {\doibase
  10.1103/PhysRevX.7.031016} {\bibfield  {journal} {\bibinfo  {journal} {Phys.
  Rev. X}\ }\textbf {\bibinfo {volume} {7}},\ \bibinfo {pages} {031016}
  (\bibinfo {year} {2017})}\BibitemShut {NoStop}%
\bibitem [{\citenamefont {Corwin}(2011)}]{2011arxivkpz}%
  \BibitemOpen
  \bibfield  {author} {\bibinfo {author} {\bibfnamefont {I.}~\bibnamefont
  {Corwin}},\ }\href {https://arxiv.org/abs/1106.1596} {\enquote {\bibinfo
  {title} {The kardar-parisi-zhang equation and universality class},}\ }
  (\bibinfo {year} {2011}),\ \Eprint {http://arxiv.org/abs/1106.1596}
  {arXiv:1106.1596 [math.PR]} \BibitemShut {NoStop}%
\bibitem [{\citenamefont {Lambek}(2007)}]{2007pregroup}%
  \BibitemOpen
  \bibfield  {author} {\bibinfo {author} {\bibfnamefont {J.}~\bibnamefont
  {Lambek}},\ }\href {\doibase 10.1007/s10849-006-9035-9} {\bibfield  {journal}
  {\bibinfo  {journal} {Journal of Logic, Language and Information}\ }\textbf
  {\bibinfo {volume} {16}},\ \bibinfo {pages} {303} (\bibinfo {year}
  {2007})}\BibitemShut {NoStop}%
\bibitem [{\citenamefont {Coecke}\ \emph {et~al.}(2010)\citenamefont {Coecke},
  \citenamefont {Sadrzadeh},\ and\ \citenamefont {Clark}}]{2010arxivpregroup}%
  \BibitemOpen
  \bibfield  {author} {\bibinfo {author} {\bibfnamefont {B.}~\bibnamefont
  {Coecke}}, \bibinfo {author} {\bibfnamefont {M.}~\bibnamefont {Sadrzadeh}}, \
  and\ \bibinfo {author} {\bibfnamefont {S.}~\bibnamefont {Clark}},\ }\href
  {https://arxiv.org/abs/1003.4394} {\enquote {\bibinfo {title} {Mathematical
  foundations for a compositional distributional model of meaning},}\ }
  (\bibinfo {year} {2010}),\ \Eprint {http://arxiv.org/abs/1003.4394}
  {arXiv:1003.4394 [cs.CL]} \BibitemShut {NoStop}%
\bibitem [{\citenamefont {Zeng}\ and\ \citenamefont
  {Coecke}(2016)}]{2016arxivpregroup}%
  \BibitemOpen
  \bibfield  {author} {\bibinfo {author} {\bibfnamefont {W.}~\bibnamefont
  {Zeng}}\ and\ \bibinfo {author} {\bibfnamefont {B.}~\bibnamefont {Coecke}},\
  }\href {\doibase 10.4204/eptcs.221.8} {\bibfield  {journal} {\bibinfo
  {journal} {Electronic Proceedings in Theoretical Computer Science}\ }\textbf
  {\bibinfo {volume} {221}},\ \bibinfo {pages} {67–75} (\bibinfo {year}
  {2016})}\BibitemShut {NoStop}%
\bibitem [{\citenamefont {Meichanetzidis}\ \emph {et~al.}(2023)\citenamefont
  {Meichanetzidis}, \citenamefont {Toumi}, \citenamefont {de~Felice},\ and\
  \citenamefont {Coecke}}]{2023qmipregroup}%
  \BibitemOpen
  \bibfield  {author} {\bibinfo {author} {\bibfnamefont {K.}~\bibnamefont
  {Meichanetzidis}}, \bibinfo {author} {\bibfnamefont {A.}~\bibnamefont
  {Toumi}}, \bibinfo {author} {\bibfnamefont {G.}~\bibnamefont {de~Felice}}, \
  and\ \bibinfo {author} {\bibfnamefont {B.}~\bibnamefont {Coecke}},\ }\href
  {\doibase 10.1007/s42484-023-00097-1} {\bibfield  {journal} {\bibinfo
  {journal} {Quantum Machine Intelligence}\ }\textbf {\bibinfo {volume} {5}},\
  \bibinfo {pages} {10} (\bibinfo {year} {2023})}\BibitemShut {NoStop}%
\bibitem [{\citenamefont {Affleck}\ \emph {et~al.}(1987)\citenamefont
  {Affleck}, \citenamefont {Kennedy}, \citenamefont {Lieb},\ and\ \citenamefont
  {Tasaki}}]{1987prlaklt}%
  \BibitemOpen
  \bibfield  {author} {\bibinfo {author} {\bibfnamefont {I.}~\bibnamefont
  {Affleck}}, \bibinfo {author} {\bibfnamefont {T.}~\bibnamefont {Kennedy}},
  \bibinfo {author} {\bibfnamefont {E.~H.}\ \bibnamefont {Lieb}}, \ and\
  \bibinfo {author} {\bibfnamefont {H.}~\bibnamefont {Tasaki}},\ }\href
  {\doibase 10.1103/PhysRevLett.59.799} {\bibfield  {journal} {\bibinfo
  {journal} {Phys. Rev. Lett.}\ }\textbf {\bibinfo {volume} {59}},\ \bibinfo
  {pages} {799} (\bibinfo {year} {1987})}\BibitemShut {NoStop}%
\bibitem [{\citenamefont {Affleck}\ \emph {et~al.}(1988)\citenamefont
  {Affleck}, \citenamefont {Kennedy}, \citenamefont {Lieb},\ and\ \citenamefont
  {Tasaki}}]{1988cmpaklt}%
  \BibitemOpen
  \bibfield  {author} {\bibinfo {author} {\bibfnamefont {I.}~\bibnamefont
  {Affleck}}, \bibinfo {author} {\bibfnamefont {T.}~\bibnamefont {Kennedy}},
  \bibinfo {author} {\bibfnamefont {E.~H.}\ \bibnamefont {Lieb}}, \ and\
  \bibinfo {author} {\bibfnamefont {H.}~\bibnamefont {Tasaki}},\ }\href
  {\doibase 10.1007/BF01218021} {\bibfield  {journal} {\bibinfo  {journal}
  {Communications in Mathematical Physics}\ }\textbf {\bibinfo {volume}
  {115}},\ \bibinfo {pages} {477} (\bibinfo {year} {1988})}\BibitemShut
  {NoStop}%
\bibitem [{\citenamefont {Moudgalya}\ \emph {et~al.}(2018)\citenamefont
  {Moudgalya}, \citenamefont {Rachel}, \citenamefont {Bernevig},\ and\
  \citenamefont {Regnault}}]{2018prbaklt}%
  \BibitemOpen
  \bibfield  {author} {\bibinfo {author} {\bibfnamefont {S.}~\bibnamefont
  {Moudgalya}}, \bibinfo {author} {\bibfnamefont {S.}~\bibnamefont {Rachel}},
  \bibinfo {author} {\bibfnamefont {B.~A.}\ \bibnamefont {Bernevig}}, \ and\
  \bibinfo {author} {\bibfnamefont {N.}~\bibnamefont {Regnault}},\ }\href
  {\doibase 10.1103/PhysRevB.98.235155} {\bibfield  {journal} {\bibinfo
  {journal} {Phys. Rev. B}\ }\textbf {\bibinfo {volume} {98}},\ \bibinfo
  {pages} {235155} (\bibinfo {year} {2018})}\BibitemShut {NoStop}%
\bibitem [{\citenamefont {Fagotti}\ and\ \citenamefont
  {Essler}(2013)}]{2013prbdistance}%
  \BibitemOpen
  \bibfield  {author} {\bibinfo {author} {\bibfnamefont {M.}~\bibnamefont
  {Fagotti}}\ and\ \bibinfo {author} {\bibfnamefont {F.~H.~L.}\ \bibnamefont
  {Essler}},\ }\href {\doibase 10.1103/PhysRevB.87.245107} {\bibfield
  {journal} {\bibinfo  {journal} {Phys. Rev. B}\ }\textbf {\bibinfo {volume}
  {87}},\ \bibinfo {pages} {245107} (\bibinfo {year} {2013})}\BibitemShut
  {NoStop}%
\bibitem [{\citenamefont {Farrelly}(2020)}]{2020quantumca}%
  \BibitemOpen
  \bibfield  {author} {\bibinfo {author} {\bibfnamefont {T.}~\bibnamefont
  {Farrelly}},\ }\href {\doibase 10.22331/q-2020-11-30-368} {\bibfield
  {journal} {\bibinfo  {journal} {{Quantum}}\ }\textbf {\bibinfo {volume}
  {4}},\ \bibinfo {pages} {368} (\bibinfo {year} {2020})}\BibitemShut {NoStop}%
\bibitem [{\citenamefont {Pozsgay}(2021)}]{2021JPAca}%
  \BibitemOpen
  \bibfield  {author} {\bibinfo {author} {\bibfnamefont {B.}~\bibnamefont
  {Pozsgay}},\ }\href {\doibase 10.1088/1751-8121/ac1dbf} {\bibfield  {journal}
  {\bibinfo  {journal} {Journal of Physics A: Mathematical and Theoretical}\
  }\textbf {\bibinfo {volume} {54}},\ \bibinfo {pages} {384001} (\bibinfo
  {year} {2021})}\BibitemShut {NoStop}%
\bibitem [{\citenamefont {Brighi}\ \emph {et~al.}(2023)\citenamefont {Brighi},
  \citenamefont {Ljubotina},\ and\ \citenamefont {Serbyn}}]{2023scipostca}%
  \BibitemOpen
  \bibfield  {author} {\bibinfo {author} {\bibfnamefont {P.}~\bibnamefont
  {Brighi}}, \bibinfo {author} {\bibfnamefont {M.}~\bibnamefont {Ljubotina}}, \
  and\ \bibinfo {author} {\bibfnamefont {M.}~\bibnamefont {Serbyn}},\ }\href
  {\doibase 10.21468/SciPostPhys.15.3.093} {\bibfield  {journal} {\bibinfo
  {journal} {SciPost Phys.}\ }\textbf {\bibinfo {volume} {15}},\ \bibinfo
  {pages} {093} (\bibinfo {year} {2023})}\BibitemShut {NoStop}%
\bibitem [{\citenamefont {Khan}\ and\ \citenamefont
  {Perkowski}(2007)}]{2007jsaqutrit}%
  \BibitemOpen
  \bibfield  {author} {\bibinfo {author} {\bibfnamefont {M.~H.}\ \bibnamefont
  {Khan}}\ and\ \bibinfo {author} {\bibfnamefont {M.~A.}\ \bibnamefont
  {Perkowski}},\ }\href {\doibase https://doi.org/10.1016/j.sysarc.2007.01.007}
  {\bibfield  {journal} {\bibinfo  {journal} {Journal of Systems Architecture}\
  }\textbf {\bibinfo {volume} {53}},\ \bibinfo {pages} {453} (\bibinfo {year}
  {2007})}\BibitemShut {NoStop}%
\bibitem [{\citenamefont {Haghparast}\ \emph {et~al.}(2017)\citenamefont
  {Haghparast}, \citenamefont {Wille},\ and\ \citenamefont
  {Monfared}}]{2017qipqutrit}%
  \BibitemOpen
  \bibfield  {author} {\bibinfo {author} {\bibfnamefont {M.}~\bibnamefont
  {Haghparast}}, \bibinfo {author} {\bibfnamefont {R.}~\bibnamefont {Wille}}, \
  and\ \bibinfo {author} {\bibfnamefont {A.~T.}\ \bibnamefont {Monfared}},\
  }\href {\doibase 10.1007/s11128-017-1735-3} {\bibfield  {journal} {\bibinfo
  {journal} {Quantum Information Processing}\ }\textbf {\bibinfo {volume}
  {16}},\ \bibinfo {pages} {284} (\bibinfo {year} {2017})}\BibitemShut
  {NoStop}%
\bibitem [{\citenamefont {Walter}\ and\ \citenamefont
  {Barkema}(2015)}]{2015paMonteCarlo}%
  \BibitemOpen
  \bibfield  {author} {\bibinfo {author} {\bibfnamefont {J.-C.}\ \bibnamefont
  {Walter}}\ and\ \bibinfo {author} {\bibfnamefont {G.}~\bibnamefont
  {Barkema}},\ }\href {\doibase https://doi.org/10.1016/j.physa.2014.06.014}
  {\bibfield  {journal} {\bibinfo  {journal} {Physica A: Statistical Mechanics
  and its Applications}\ }\textbf {\bibinfo {volume} {418}},\ \bibinfo {pages}
  {78} (\bibinfo {year} {2015})},\ \bibinfo {note} {proceedings of the 13th
  International Summer School on Fundamental Problems in Statistical
  Physics}\BibitemShut {NoStop}%
\bibitem [{\citenamefont {Ma}\ \emph {et~al.}(2022)\citenamefont {Ma},
  \citenamefont {Han}, \citenamefont {Meir},\ and\ \citenamefont
  {Sela}}]{2022praea}%
  \BibitemOpen
  \bibfield  {author} {\bibinfo {author} {\bibfnamefont {Z.}~\bibnamefont
  {Ma}}, \bibinfo {author} {\bibfnamefont {C.}~\bibnamefont {Han}}, \bibinfo
  {author} {\bibfnamefont {Y.}~\bibnamefont {Meir}}, \ and\ \bibinfo {author}
  {\bibfnamefont {E.}~\bibnamefont {Sela}},\ }\href {\doibase
  10.1103/PhysRevA.105.042416} {\bibfield  {journal} {\bibinfo  {journal}
  {Phys. Rev. A}\ }\textbf {\bibinfo {volume} {105}},\ \bibinfo {pages}
  {042416} (\bibinfo {year} {2022})}\BibitemShut {NoStop}%
\bibitem [{\citenamefont {Ares}\ \emph {et~al.}(2023)\citenamefont {Ares},
  \citenamefont {Murciano},\ and\ \citenamefont {Calabrese}}]{2023ncea}%
  \BibitemOpen
  \bibfield  {author} {\bibinfo {author} {\bibfnamefont {F.}~\bibnamefont
  {Ares}}, \bibinfo {author} {\bibfnamefont {S.}~\bibnamefont {Murciano}}, \
  and\ \bibinfo {author} {\bibfnamefont {P.}~\bibnamefont {Calabrese}},\ }\href
  {\doibase 10.1038/s41467-023-37747-8} {\bibfield  {journal} {\bibinfo
  {journal} {Nature Communications}\ }\textbf {\bibinfo {volume} {14}},\
  \bibinfo {pages} {2036} (\bibinfo {year} {2023})}\BibitemShut {NoStop}%
\bibitem [{\citenamefont {Khor}\ \emph {et~al.}(2024)\citenamefont {Khor},
  \citenamefont {K{\"{u}}rk{\c{c}}{\"{u}}oglu}, \citenamefont {Hobbs},
  \citenamefont {Perdue},\ and\ \citenamefont {Klich}}]{2024quantumea}%
  \BibitemOpen
  \bibfield  {author} {\bibinfo {author} {\bibfnamefont {B.~J.~J.}\
  \bibnamefont {Khor}}, \bibinfo {author} {\bibfnamefont {D.~M.}\ \bibnamefont
  {K{\"{u}}rk{\c{c}}{\"{u}}oglu}}, \bibinfo {author} {\bibfnamefont {T.~J.}\
  \bibnamefont {Hobbs}}, \bibinfo {author} {\bibfnamefont {G.~N.}\ \bibnamefont
  {Perdue}}, \ and\ \bibinfo {author} {\bibfnamefont {I.}~\bibnamefont
  {Klich}},\ }\href {\doibase 10.22331/q-2024-09-06-1462} {\bibfield  {journal}
  {\bibinfo  {journal} {{Quantum}}\ }\textbf {\bibinfo {volume} {8}},\ \bibinfo
  {pages} {1462} (\bibinfo {year} {2024})}\BibitemShut {NoStop}%
\bibitem [{\citenamefont {Joshi}\ \emph {et~al.}(2024)\citenamefont {Joshi},
  \citenamefont {Franke}, \citenamefont {Rath}, \citenamefont {Ares},
  \citenamefont {Murciano}, \citenamefont {Kranzl}, \citenamefont {Blatt},
  \citenamefont {Zoller}, \citenamefont {Vermersch}, \citenamefont {Calabrese},
  \citenamefont {Roos},\ and\ \citenamefont {Joshi}}]{2024prlea}%
  \BibitemOpen
  \bibfield  {author} {\bibinfo {author} {\bibfnamefont {L.~K.}\ \bibnamefont
  {Joshi}}, \bibinfo {author} {\bibfnamefont {J.}~\bibnamefont {Franke}},
  \bibinfo {author} {\bibfnamefont {A.}~\bibnamefont {Rath}}, \bibinfo {author}
  {\bibfnamefont {F.}~\bibnamefont {Ares}}, \bibinfo {author} {\bibfnamefont
  {S.}~\bibnamefont {Murciano}}, \bibinfo {author} {\bibfnamefont
  {F.}~\bibnamefont {Kranzl}}, \bibinfo {author} {\bibfnamefont
  {R.}~\bibnamefont {Blatt}}, \bibinfo {author} {\bibfnamefont
  {P.}~\bibnamefont {Zoller}}, \bibinfo {author} {\bibfnamefont
  {B.}~\bibnamefont {Vermersch}}, \bibinfo {author} {\bibfnamefont
  {P.}~\bibnamefont {Calabrese}}, \bibinfo {author} {\bibfnamefont {C.~F.}\
  \bibnamefont {Roos}}, \ and\ \bibinfo {author} {\bibfnamefont {M.~K.}\
  \bibnamefont {Joshi}},\ }\href {\doibase 10.1103/PhysRevLett.133.010402}
  {\bibfield  {journal} {\bibinfo  {journal} {Phys. Rev. Lett.}\ }\textbf
  {\bibinfo {volume} {133}},\ \bibinfo {pages} {010402} (\bibinfo {year}
  {2024})}\BibitemShut {NoStop}%
\bibitem [{\citenamefont {Sala}\ \emph {et~al.}(2020)\citenamefont {Sala},
  \citenamefont {Rakovszky}, \citenamefont {Verresen}, \citenamefont {Knap},\
  and\ \citenamefont {Pollmann}}]{2020prxfrag}%
  \BibitemOpen
  \bibfield  {author} {\bibinfo {author} {\bibfnamefont {P.}~\bibnamefont
  {Sala}}, \bibinfo {author} {\bibfnamefont {T.}~\bibnamefont {Rakovszky}},
  \bibinfo {author} {\bibfnamefont {R.}~\bibnamefont {Verresen}}, \bibinfo
  {author} {\bibfnamefont {M.}~\bibnamefont {Knap}}, \ and\ \bibinfo {author}
  {\bibfnamefont {F.}~\bibnamefont {Pollmann}},\ }\href {\doibase
  10.1103/PhysRevX.10.011047} {\bibfield  {journal} {\bibinfo  {journal} {Phys.
  Rev. X}\ }\textbf {\bibinfo {volume} {10}},\ \bibinfo {pages} {011047}
  (\bibinfo {year} {2020})}\BibitemShut {NoStop}%
\bibitem [{\citenamefont {Feng}\ and\ \citenamefont
  {Skinner}(2022)}]{2022prrfrag}%
  \BibitemOpen
  \bibfield  {author} {\bibinfo {author} {\bibfnamefont {X.}~\bibnamefont
  {Feng}}\ and\ \bibinfo {author} {\bibfnamefont {B.}~\bibnamefont {Skinner}},\
  }\href {\doibase 10.1103/PhysRevResearch.4.013053} {\bibfield  {journal}
  {\bibinfo  {journal} {Phys. Rev. Res.}\ }\textbf {\bibinfo {volume} {4}},\
  \bibinfo {pages} {013053} (\bibinfo {year} {2022})}\BibitemShut {NoStop}%
\bibitem [{\citenamefont {Moudgalya}\ \emph {et~al.}(2022)\citenamefont
  {Moudgalya}, \citenamefont {Bernevig},\ and\ \citenamefont
  {Regnault}}]{2022rppfrag}%
  \BibitemOpen
  \bibfield  {author} {\bibinfo {author} {\bibfnamefont {S.}~\bibnamefont
  {Moudgalya}}, \bibinfo {author} {\bibfnamefont {B.~A.}\ \bibnamefont
  {Bernevig}}, \ and\ \bibinfo {author} {\bibfnamefont {N.}~\bibnamefont
  {Regnault}},\ }\href {\doibase 10.1088/1361-6633/ac73a0} {\bibfield
  {journal} {\bibinfo  {journal} {Reports on Progress in Physics}\ }\textbf
  {\bibinfo {volume} {85}},\ \bibinfo {pages} {086501} (\bibinfo {year}
  {2022})}\BibitemShut {NoStop}%
\bibitem [{\citenamefont {Kesten}(1980)}]{1980cmppercolation}%
  \BibitemOpen
  \bibfield  {author} {\bibinfo {author} {\bibfnamefont {H.}~\bibnamefont
  {Kesten}},\ }\href {\doibase 10.1007/BF01197577} {\bibfield  {journal}
  {\bibinfo  {journal} {Communications in Mathematical Physics}\ }\textbf
  {\bibinfo {volume} {74}},\ \bibinfo {pages} {41} (\bibinfo {year}
  {1980})}\BibitemShut {NoStop}%
\bibitem [{\citenamefont {Saberi}(2015)}]{2015prpercolation}%
  \BibitemOpen
  \bibfield  {author} {\bibinfo {author} {\bibfnamefont {A.~A.}\ \bibnamefont
  {Saberi}},\ }\href {\doibase https://doi.org/10.1016/j.physrep.2015.03.003}
  {\bibfield  {journal} {\bibinfo  {journal} {Physics Reports}\ }\textbf
  {\bibinfo {volume} {578}},\ \bibinfo {pages} {1} (\bibinfo {year} {2015})},\
  \bibinfo {note} {recent advances in percolation theory and its
  applications}\BibitemShut {NoStop}%
\bibitem [{\citenamefont {Meng}\ \emph {et~al.}(2021)\citenamefont {Meng},
  \citenamefont {Gao},\ and\ \citenamefont {Havlin}}]{2021prlpercolation}%
  \BibitemOpen
  \bibfield  {author} {\bibinfo {author} {\bibfnamefont {X.}~\bibnamefont
  {Meng}}, \bibinfo {author} {\bibfnamefont {J.}~\bibnamefont {Gao}}, \ and\
  \bibinfo {author} {\bibfnamefont {S.}~\bibnamefont {Havlin}},\ }\href
  {\doibase 10.1103/PhysRevLett.126.170501} {\bibfield  {journal} {\bibinfo
  {journal} {Phys. Rev. Lett.}\ }\textbf {\bibinfo {volume} {126}},\ \bibinfo
  {pages} {170501} (\bibinfo {year} {2021})}\BibitemShut {NoStop}%
\bibitem [{\citenamefont {Rota}(2016)}]{2016jheptmi}%
  \BibitemOpen
  \bibfield  {author} {\bibinfo {author} {\bibfnamefont {M.}~\bibnamefont
  {Rota}},\ }\href {\doibase 10.1007/JHEP04(2016)075} {\bibfield  {journal}
  {\bibinfo  {journal} {Journal of High Energy Physics}\ }\textbf {\bibinfo
  {volume} {2016}},\ \bibinfo {pages} {75} (\bibinfo {year}
  {2016})}\BibitemShut {NoStop}%
\bibitem [{\citenamefont {Hosur}\ \emph {et~al.}(2016)\citenamefont {Hosur},
  \citenamefont {Qi}, \citenamefont {Roberts},\ and\ \citenamefont
  {Yoshida}}]{2016jhepchaos}%
  \BibitemOpen
  \bibfield  {author} {\bibinfo {author} {\bibfnamefont {P.}~\bibnamefont
  {Hosur}}, \bibinfo {author} {\bibfnamefont {X.-L.}\ \bibnamefont {Qi}},
  \bibinfo {author} {\bibfnamefont {D.~A.}\ \bibnamefont {Roberts}}, \ and\
  \bibinfo {author} {\bibfnamefont {B.}~\bibnamefont {Yoshida}},\ }\href
  {\doibase 10.1007/JHEP02(2016)004} {\bibfield  {journal} {\bibinfo  {journal}
  {Journal of High Energy Physics}\ }\textbf {\bibinfo {volume} {2016}},\
  \bibinfo {pages} {4} (\bibinfo {year} {2016})}\BibitemShut {NoStop}%
\bibitem [{\citenamefont {Seshadri}\ \emph {et~al.}(2018)\citenamefont
  {Seshadri}, \citenamefont {Madhok},\ and\ \citenamefont
  {Lakshminarayan}}]{2018pretmi}%
  \BibitemOpen
  \bibfield  {author} {\bibinfo {author} {\bibfnamefont {A.}~\bibnamefont
  {Seshadri}}, \bibinfo {author} {\bibfnamefont {V.}~\bibnamefont {Madhok}}, \
  and\ \bibinfo {author} {\bibfnamefont {A.}~\bibnamefont {Lakshminarayan}},\
  }\href {\doibase 10.1103/PhysRevE.98.052205} {\bibfield  {journal} {\bibinfo
  {journal} {Phys. Rev. E}\ }\textbf {\bibinfo {volume} {98}},\ \bibinfo
  {pages} {052205} (\bibinfo {year} {2018})}\BibitemShut {NoStop}%
\bibitem [{\citenamefont {Kitaev}\ and\ \citenamefont
  {Preskill}(2006)}]{2006prlTopological}%
  \BibitemOpen
  \bibfield  {author} {\bibinfo {author} {\bibfnamefont {A.}~\bibnamefont
  {Kitaev}}\ and\ \bibinfo {author} {\bibfnamefont {J.}~\bibnamefont
  {Preskill}},\ }\href {\doibase 10.1103/PhysRevLett.96.110404} {\bibfield
  {journal} {\bibinfo  {journal} {Phys. Rev. Lett.}\ }\textbf {\bibinfo
  {volume} {96}},\ \bibinfo {pages} {110404} (\bibinfo {year}
  {2006})}\BibitemShut {NoStop}%
\bibitem [{\citenamefont {Iyoda}\ and\ \citenamefont
  {Sagawa}(2018)}]{2018prascrambling}%
  \BibitemOpen
  \bibfield  {author} {\bibinfo {author} {\bibfnamefont {E.}~\bibnamefont
  {Iyoda}}\ and\ \bibinfo {author} {\bibfnamefont {T.}~\bibnamefont {Sagawa}},\
  }\href {\doibase 10.1103/PhysRevA.97.042330} {\bibfield  {journal} {\bibinfo
  {journal} {Phys. Rev. A}\ }\textbf {\bibinfo {volume} {97}},\ \bibinfo
  {pages} {042330} (\bibinfo {year} {2018})}\BibitemShut {NoStop}%
\bibitem [{\citenamefont {Schnaack}\ \emph {et~al.}(2019)\citenamefont
  {Schnaack}, \citenamefont {B\"olter}, \citenamefont {Paeckel}, \citenamefont
  {Manmana}, \citenamefont {Kehrein},\ and\ \citenamefont
  {Schmitt}}]{2019prbscrambling}%
  \BibitemOpen
  \bibfield  {author} {\bibinfo {author} {\bibfnamefont {O.}~\bibnamefont
  {Schnaack}}, \bibinfo {author} {\bibfnamefont {N.}~\bibnamefont {B\"olter}},
  \bibinfo {author} {\bibfnamefont {S.}~\bibnamefont {Paeckel}}, \bibinfo
  {author} {\bibfnamefont {S.~R.}\ \bibnamefont {Manmana}}, \bibinfo {author}
  {\bibfnamefont {S.}~\bibnamefont {Kehrein}}, \ and\ \bibinfo {author}
  {\bibfnamefont {M.}~\bibnamefont {Schmitt}},\ }\href {\doibase
  10.1103/PhysRevB.100.224302} {\bibfield  {journal} {\bibinfo  {journal}
  {Phys. Rev. B}\ }\textbf {\bibinfo {volume} {100}},\ \bibinfo {pages}
  {224302} (\bibinfo {year} {2019})}\BibitemShut {NoStop}%
\bibitem [{\citenamefont {Xu}\ and\ \citenamefont
  {Swingle}(2024)}]{2024prxqscrambling}%
  \BibitemOpen
  \bibfield  {author} {\bibinfo {author} {\bibfnamefont {S.}~\bibnamefont
  {Xu}}\ and\ \bibinfo {author} {\bibfnamefont {B.}~\bibnamefont {Swingle}},\
  }\href {\doibase 10.1103/PRXQuantum.5.010201} {\bibfield  {journal} {\bibinfo
   {journal} {PRX Quantum}\ }\textbf {\bibinfo {volume} {5}},\ \bibinfo {pages}
  {010201} (\bibinfo {year} {2024})}\BibitemShut {NoStop}%
\bibitem [{\citenamefont {Schollwöck}(2011)}]{2011aopmps}%
  \BibitemOpen
  \bibfield  {author} {\bibinfo {author} {\bibfnamefont {U.}~\bibnamefont
  {Schollwöck}},\ }\href {\doibase https://doi.org/10.1016/j.aop.2010.09.012}
  {\bibfield  {journal} {\bibinfo  {journal} {Annals of Physics}\ }\textbf
  {\bibinfo {volume} {326}},\ \bibinfo {pages} {96} (\bibinfo {year} {2011})},\
  \bibinfo {note} {january 2011 Special Issue}\BibitemShut {NoStop}%
\bibitem [{\citenamefont {Cirac}\ \emph {et~al.}(2021)\citenamefont {Cirac},
  \citenamefont {P\'erez-Garc\'{\i}a}, \citenamefont {Schuch},\ and\
  \citenamefont {Verstraete}}]{2021rmpmps}%
  \BibitemOpen
  \bibfield  {author} {\bibinfo {author} {\bibfnamefont {J.~I.}\ \bibnamefont
  {Cirac}}, \bibinfo {author} {\bibfnamefont {D.}~\bibnamefont
  {P\'erez-Garc\'{\i}a}}, \bibinfo {author} {\bibfnamefont {N.}~\bibnamefont
  {Schuch}}, \ and\ \bibinfo {author} {\bibfnamefont {F.}~\bibnamefont
  {Verstraete}},\ }\href {\doibase 10.1103/RevModPhys.93.045003} {\bibfield
  {journal} {\bibinfo  {journal} {Rev. Mod. Phys.}\ }\textbf {\bibinfo {volume}
  {93}},\ \bibinfo {pages} {045003} (\bibinfo {year} {2021})}\BibitemShut
  {NoStop}%
\bibitem [{\citenamefont {O’Riordan}\ \emph {et~al.}(2020)\citenamefont
  {O’Riordan}, \citenamefont {Doyle}, \citenamefont {Baruffa},\ and\
  \citenamefont {Kannan}}]{2021mlstbinary}%
  \BibitemOpen
  \bibfield  {author} {\bibinfo {author} {\bibfnamefont {L.~J.}\ \bibnamefont
  {O’Riordan}}, \bibinfo {author} {\bibfnamefont {M.}~\bibnamefont {Doyle}},
  \bibinfo {author} {\bibfnamefont {F.}~\bibnamefont {Baruffa}}, \ and\
  \bibinfo {author} {\bibfnamefont {V.}~\bibnamefont {Kannan}},\ }\href
  {\doibase 10.1088/2632-2153/abbd2e} {\bibfield  {journal} {\bibinfo
  {journal} {Machine Learning: Science and Technology}\ }\textbf {\bibinfo
  {volume} {2}},\ \bibinfo {pages} {015011} (\bibinfo {year}
  {2020})}\BibitemShut {NoStop}%
\bibitem [{\citenamefont {Bernien}\ \emph {et~al.}(2017)\citenamefont
  {Bernien}, \citenamefont {Schwartz}, \citenamefont {Keesling}, \citenamefont
  {Levine}, \citenamefont {Omran}, \citenamefont {Pichler}, \citenamefont
  {Choi}, \citenamefont {Zibrov}, \citenamefont {Endres}, \citenamefont
  {Greiner}, \citenamefont {Vuleti{\'{c}}},\ and\ \citenamefont
  {Lukin}}]{2017natsimulator}%
  \BibitemOpen
  \bibfield  {author} {\bibinfo {author} {\bibfnamefont {H.}~\bibnamefont
  {Bernien}}, \bibinfo {author} {\bibfnamefont {S.}~\bibnamefont {Schwartz}},
  \bibinfo {author} {\bibfnamefont {A.}~\bibnamefont {Keesling}}, \bibinfo
  {author} {\bibfnamefont {H.}~\bibnamefont {Levine}}, \bibinfo {author}
  {\bibfnamefont {A.}~\bibnamefont {Omran}}, \bibinfo {author} {\bibfnamefont
  {H.}~\bibnamefont {Pichler}}, \bibinfo {author} {\bibfnamefont
  {S.}~\bibnamefont {Choi}}, \bibinfo {author} {\bibfnamefont {A.~S.}\
  \bibnamefont {Zibrov}}, \bibinfo {author} {\bibfnamefont {M.}~\bibnamefont
  {Endres}}, \bibinfo {author} {\bibfnamefont {M.}~\bibnamefont {Greiner}},
  \bibinfo {author} {\bibfnamefont {V.}~\bibnamefont {Vuleti{\'{c}}}}, \ and\
  \bibinfo {author} {\bibfnamefont {M.~D.}\ \bibnamefont {Lukin}},\ }\href
  {\doibase 10.1038/nature24622} {\bibfield  {journal} {\bibinfo  {journal}
  {Nature}\ }\textbf {\bibinfo {volume} {551}},\ \bibinfo {pages} {579}
  (\bibinfo {year} {2017})}\BibitemShut {NoStop}%
\bibitem [{\citenamefont {Barends}\ \emph {et~al.}(2015)\citenamefont
  {Barends}, \citenamefont {Lamata}, \citenamefont {Kelly}, \citenamefont
  {Garc{\'i}a-{\'A}lvarez}, \citenamefont {Fowler}, \citenamefont {Megrant},
  \citenamefont {Jeffrey}, \citenamefont {White}, \citenamefont {Sank},
  \citenamefont {Mutus}, \citenamefont {Campbell}, \citenamefont {Chen},
  \citenamefont {Chen}, \citenamefont {Chiaro}, \citenamefont {Dunsworth},
  \citenamefont {Hoi}, \citenamefont {Neill}, \citenamefont {O'Malley},
  \citenamefont {Quintana}, \citenamefont {Roushan}, \citenamefont
  {Vainsencher}, \citenamefont {Wenner}, \citenamefont {Solano},\ and\
  \citenamefont {Martinis}}]{2015ncsimulator}%
  \BibitemOpen
  \bibfield  {author} {\bibinfo {author} {\bibfnamefont {R.}~\bibnamefont
  {Barends}}, \bibinfo {author} {\bibfnamefont {L.}~\bibnamefont {Lamata}},
  \bibinfo {author} {\bibfnamefont {J.}~\bibnamefont {Kelly}}, \bibinfo
  {author} {\bibfnamefont {L.}~\bibnamefont {Garc{\'i}a-{\'A}lvarez}}, \bibinfo
  {author} {\bibfnamefont {A.~G.}\ \bibnamefont {Fowler}}, \bibinfo {author}
  {\bibfnamefont {A.}~\bibnamefont {Megrant}}, \bibinfo {author} {\bibfnamefont
  {E.}~\bibnamefont {Jeffrey}}, \bibinfo {author} {\bibfnamefont {T.~C.}\
  \bibnamefont {White}}, \bibinfo {author} {\bibfnamefont {D.}~\bibnamefont
  {Sank}}, \bibinfo {author} {\bibfnamefont {J.~Y.}\ \bibnamefont {Mutus}},
  \bibinfo {author} {\bibfnamefont {B.}~\bibnamefont {Campbell}}, \bibinfo
  {author} {\bibfnamefont {Y.}~\bibnamefont {Chen}}, \bibinfo {author}
  {\bibfnamefont {Z.}~\bibnamefont {Chen}}, \bibinfo {author} {\bibfnamefont
  {B.}~\bibnamefont {Chiaro}}, \bibinfo {author} {\bibfnamefont
  {A.}~\bibnamefont {Dunsworth}}, \bibinfo {author} {\bibfnamefont {I.-C.}\
  \bibnamefont {Hoi}}, \bibinfo {author} {\bibfnamefont {C.}~\bibnamefont
  {Neill}}, \bibinfo {author} {\bibfnamefont {P.~J.~J.}\ \bibnamefont
  {O'Malley}}, \bibinfo {author} {\bibfnamefont {C.}~\bibnamefont {Quintana}},
  \bibinfo {author} {\bibfnamefont {P.}~\bibnamefont {Roushan}}, \bibinfo
  {author} {\bibfnamefont {A.}~\bibnamefont {Vainsencher}}, \bibinfo {author}
  {\bibfnamefont {J.}~\bibnamefont {Wenner}}, \bibinfo {author} {\bibfnamefont
  {E.}~\bibnamefont {Solano}}, \ and\ \bibinfo {author} {\bibfnamefont {J.~M.}\
  \bibnamefont {Martinis}},\ }\href {\doibase 10.1038/ncomms8654} {\bibfield
  {journal} {\bibinfo  {journal} {Nature Communications}\ }\textbf {\bibinfo
  {volume} {6}},\ \bibinfo {pages} {7654} (\bibinfo {year} {2015})}\BibitemShut
  {NoStop}%
\bibitem [{\citenamefont {Shi}\ \emph {et~al.}(2024)\citenamefont {Shi},
  \citenamefont {Sun}, \citenamefont {Wang}, \citenamefont {Wang},
  \citenamefont {Zhang}, \citenamefont {Ma}, \citenamefont {Liu}, \citenamefont
  {Zhao}, \citenamefont {Song}, \citenamefont {Liang}, \citenamefont {Mei},
  \citenamefont {Li}, \citenamefont {Chen}, \citenamefont {Song}, \citenamefont
  {Wang}, \citenamefont {Xue}, \citenamefont {Yu}, \citenamefont {Huang},
  \citenamefont {Xiang}, \citenamefont {Xu}, \citenamefont {Zheng},\ and\
  \citenamefont {Fan}}]{2024ncsimulator}%
  \BibitemOpen
  \bibfield  {author} {\bibinfo {author} {\bibfnamefont {Y.-H.}\ \bibnamefont
  {Shi}}, \bibinfo {author} {\bibfnamefont {Z.-H.}\ \bibnamefont {Sun}},
  \bibinfo {author} {\bibfnamefont {Y.-Y.}\ \bibnamefont {Wang}}, \bibinfo
  {author} {\bibfnamefont {Z.-A.}\ \bibnamefont {Wang}}, \bibinfo {author}
  {\bibfnamefont {Y.-R.}\ \bibnamefont {Zhang}}, \bibinfo {author}
  {\bibfnamefont {W.-G.}\ \bibnamefont {Ma}}, \bibinfo {author} {\bibfnamefont
  {H.-T.}\ \bibnamefont {Liu}}, \bibinfo {author} {\bibfnamefont
  {K.}~\bibnamefont {Zhao}}, \bibinfo {author} {\bibfnamefont {J.-C.}\
  \bibnamefont {Song}}, \bibinfo {author} {\bibfnamefont {G.-H.}\ \bibnamefont
  {Liang}}, \bibinfo {author} {\bibfnamefont {J.-C.}\ \bibnamefont {Mei},
  \bibfnamefont {Zheng-Yangand~Zhang}}, \bibinfo {author} {\bibfnamefont
  {H.}~\bibnamefont {Li}}, \bibinfo {author} {\bibfnamefont {C.-T.}\
  \bibnamefont {Chen}}, \bibinfo {author} {\bibfnamefont {X.}~\bibnamefont
  {Song}}, \bibinfo {author} {\bibfnamefont {J.}~\bibnamefont {Wang}}, \bibinfo
  {author} {\bibfnamefont {G.}~\bibnamefont {Xue}}, \bibinfo {author}
  {\bibfnamefont {H.}~\bibnamefont {Yu}}, \bibinfo {author} {\bibfnamefont
  {K.}~\bibnamefont {Huang}}, \bibinfo {author} {\bibfnamefont
  {Z.}~\bibnamefont {Xiang}}, \bibinfo {author} {\bibfnamefont
  {K.}~\bibnamefont {Xu}}, \bibinfo {author} {\bibfnamefont {D.}~\bibnamefont
  {Zheng}}, \ and\ \bibinfo {author} {\bibfnamefont {H.}~\bibnamefont {Fan}},\
  }\href {\doibase 10.1038/s41467-024-52082-2} {\bibfield  {journal} {\bibinfo
  {journal} {Nature Communications}\ }\textbf {\bibinfo {volume} {15}},\
  \bibinfo {pages} {7573} (\bibinfo {year} {2024})}\BibitemShut {NoStop}%
\bibitem [{\citenamefont {Melko}\ and\ \citenamefont
  {Carrasquilla}(2024)}]{2024ncssimulator}%
  \BibitemOpen
  \bibfield  {author} {\bibinfo {author} {\bibfnamefont {R.~G.}\ \bibnamefont
  {Melko}}\ and\ \bibinfo {author} {\bibfnamefont {J.}~\bibnamefont
  {Carrasquilla}},\ }\href {\doibase 10.1038/s43588-023-00578-0} {\bibfield
  {journal} {\bibinfo  {journal} {Nature Computational Science}\ }\textbf
  {\bibinfo {volume} {4}},\ \bibinfo {pages} {11} (\bibinfo {year}
  {2024})}\BibitemShut {NoStop}%
\end{thebibliography}%
%\bibliography{ref}

\end{document}